\documentclass[12pt]{article}
\usepackage{myart}

\normalbaselines
\input tcilatex.tex

\begin{document}

\author{Yuri A. Rylov}
\title{Classical description of pair production. }
\date{Institute for Problems in Mechanics, Russian Academy of Sciences \\
101-1,Vernadskii Ave., Moscow, 119526, Russia \\
email: rylov@ipmnet.ru\\
Web site: {$http://rsfq1.physics.sunysb.edu/\symbol{126}rylov/yrylov.htm$}\\
or mirror Web site: {$http://gasdyn-ipm.ipmnet.ru/\symbol{126}%
rylov/yrylov.htm$}}
\maketitle

\begin{abstract}
It is shown that the classical description of pair production effect is
possible, i.e. one can describe pair production without a reference to
quantum principles. Pair production appears at statistical description of
stochastic relativistic particles. There is a special force field which is
responsible for pair production. This field is a reasonable consequence of
quantum stochasticity. Consideration of quantum systems as stochastic
systems and statistical description of them generates hydrodynamic
interpretation of quantum phenomena. In the collision problem the
hydrodynamic interpretation appears to be alternative to the conventional
interpretation, based on quantum principles and on consideration of the wave
function as a principal object of dynamics. Hydrodynamic interpretation
leads to such a statement of the collision problem, which is alternative to
the conventional S-matrix theory.
\end{abstract}

\section{Introduction}

Pair production is a specific quantum effect. It is described conventionally
in terms of quantum theory. We do not know how to describe the pair
production effects in classical terms. But if the classical description of
the pair production were possible, it would be very useful. Indeed, at high
energies of particles their de Broglie wave length $\lambda _{\mathrm{B}%
}=\hbar /p$ becomes to be very small, and description of high energy
particles becomes semiclassical or even classical. The classical description
is simpler, than the quantum one, because it uses less information and
admits a simpler interpretation. Unfortunately, we cannot realize advantages
of classical description, because at high energies the pair production
processes are dominating, but we do not know how to describe them in
classical terms. In general, we know about the pair production only that it
appears at collision of high energy particles, and it is described only in
terms of quantum theory. But what agents are responsible for the pair
production remains to be seen.

In this paper we are going to investigate what agents are responsible for
the pair production and to show that a classical description of pair
production effects is possible. Speaking about classical description, we
mean only that \textit{principles of quantum mechanics are not used} at such
a description. Classical description does not mean a classical approximation
of quantum description. Our classical description is an exact description.
In particular, it describes quantum effects, using quantum constant $\hbar $%
, which appears as a constant describing the space-time properties (but not
as an attribute of quantum principles).

In our consideration we use model conceptions of quantum phenomena (MCQP),
which is a new stage in investigation of microcosm. MCQP relates to
conventional quantum theory (axiomatic conception of quantum phenomena
(ACQP)) in the same way, as the statistical physics relates to
thermodynamics. The main difference between MCQP and ACQP in investigation
methods which are used. Methods of MCQP are more subtle and effective, than
those of ACQP. One can see this from the next table

\noindent $
\begin{array}{cc}
\text{ACQP} & \text{MCQP} \\ 
\begin{array}{c}
\text{Combination of nonrelativistic } \\ 
\text{quantum technique with } \\ 
\text{principles of relativity}
\end{array}
& 
\begin{array}{c}
\text{Consequent relativistic description } \\ 
\text{at all stages}
\end{array}
\\ 
\begin{array}{c}
\text{1. Quantization: procedure on } \\ 
\text{the conceptual level:} \\ 
\mathbf{p}\rightarrow -i\hbar \mathbf{\nabla }\;\;\;\text{etc. }
\end{array}
& 
\begin{array}{c}
\text{1. Dynamic quantization: relativistic } \\ 
\text{procedure on the dynamic level} \\ 
m^{2}\rightarrow m_{\mathrm{eff}}^{2}=m^{2}+\frac{\hbar ^{2}}{c^{2}}\left(
\kappa _{l}\kappa ^{l}+\partial _{l}\kappa ^{l}\right)
\end{array}
\\ 
\begin{array}{c}
\text{2. Transition to classical description:} \\ 
\text{procedure on conceptual level} \\ 
\hbar \rightarrow 0\qquad \psi \rightarrow \left( x,p\right)
\end{array}
& 
\begin{array}{c}
\text{2. Dynamic disquantization: relativistic} \\ 
\text{ procedure on dynamic level} \\ 
\partial ^{k}\rightarrow \frac{j^{k}j^{l}}{j_{s}j^{s}}\partial _{l}
\end{array}
\\ 
\begin{array}{c}
\text{3. Interpretation in terms of wave} \\ 
\text{function }\psi
\end{array}
& 
\begin{array}{c}
\text{3. Interpretation in terms of statistical} \\ 
\text{average world lines (WL)} \\ 
\frac{dx^{i}}{d\tau }=j^{i}\left( x\right) ,\;\;\; \\ 
j^{k}=-\frac{i\hbar }{2}\left( \psi ^{\ast }\partial ^{k}\psi -\partial
^{k}\psi ^{\ast }\cdot \psi \right)
\end{array}
\\ 
\begin{array}{c}
\text{4. One kind of measurement, as } \\ 
\text{far as only one statistical average } \\ 
\text{object }\left\langle \mathcal{S}\right\rangle \text{ is considered. It
is } \\ 
\text{referred to as quantum system}
\end{array}
& 
\begin{array}{c}
\text{4.Two kinds of measurement, because } \\ 
\text{two kinds of objects (individual }\mathcal{S}_{\mathrm{st}}\text{ } \\ 
\text{and statistical average }\left\langle \mathcal{S}\right\rangle \text{)
are } \\ 
\text{considered}
\end{array}
\\ 
\begin{array}{c}
\text{5. Additional hypotheses are used} \\ 
\text{(QM principles) }
\end{array}
& 
\begin{array}{c}
\text{5. \textit{No additional hypotheses are used}}
\end{array}
\end{array}
$

In this paper we develop technique of MCQP and apply methods of MCQP for
investigation of the pair production process. This technique is based on
consideration of quantum mechanics as a special case of the stochastic
system dynamics \cite{R2002}. The stochastic system dynamics is suitable for
description of any stochastic systems (not only quantum), and because of
this it cannot use quantum principles and does not use them. All description
is produced in classical (non-quantum) terms, although description in terms
of wave function is also possible as a special case of a description. Such
an approach is very convenient, because it gives a very simple
interpretation of quantum effects in terms of classical dynamics of
particles and fluids.

The main distinction between the quantum mechanics and dynamics of
stochastic systems lies in their relation to the wave function. In the
quantum mechanics the wave function $\psi $ is a fundamental object, whose
properties are defined by quantum principles. As a result the interpretation
of all objects which is constructed on the base of the wave function is
rather obscure, because the meaning of the wave function itself is obscure.
In the stochastic system dynamics the wave function is simply a method of
description of any nondissipative fluid \cite{R99}. In this case the meaning
of all objects constructed on the base of the wave function and the meaning
of the wave function itself are obtained rather simply in terms of classical
description of a fluid (continuous medium).

Connection between the fluid and the Schr\"{o}dinger equation is known since
the beginning of the quantum mechanics construction \cite{M26,B26}. In after
years many authors developed this interplay known as hydrodynamic
interpretation of quantum mechanics \cite
{B52,T52,T53,JZ63,JZ64,HZ69,B73,BH89,H93}. But this interpretation was
founded ultimately on the wave function as a fundamental object of dynamics.
It cannot go outside the framework of quantum principles, because the
connection between the hydrodynamic interpretation and the quantum mechanics
was one-way connection. One could obtain the irrotational fluid flow from
the dynamic equation for the wave function (Schr\"{o}dinger equation), but
one did not know how to transform dynamic equations for a fluid to the
dynamic equation for a wave function. In other words, we did not know how to
describe rotational fluid flow in terms of the wave function. In terms of
the wave function we could describe only irrotational fluid flow.

To describe arbitrary fluid flow in terms of a wave function, one needs to
integrate conventional dynamic equations for a fluid (Euler equations).
Indeed, the Schr\"{o}dinger equation 
\begin{equation}
i\hbar \frac{\partial \psi }{\partial t}+\frac{\hbar ^{2}}{2m}\mathbf{\nabla 
}^{2}\psi =0  \label{a0.1}
\end{equation}
may be reduced to the hydrodynamic equations for the variables $\rho ,%
\mathbf{v} $, describing the fluid state. Substituting $\psi =\sqrt{\rho }%
\exp \left( i\hbar \varphi \right) $ in (\ref{a0.1}) and separating real and
imaginary parts of the equation, we obtain expressions for time derivatives $%
\partial _{0}\rho $ and $\partial _{0}\varphi $. To obtain expression for
the time derivative $\partial _{0}\mathbf{v}$ of the velocity $\mathbf{v=}%
\frac{\hbar }{m}\mathbf{\nabla }\varphi $, we need to differentiate dynamic
equation for $\partial _{0}\varphi $, forming combination $\partial _{0}%
\mathbf{v=\nabla }\left( \frac{\hbar }{m}\partial _{0}\varphi \right) $. The
reverse transition from hydrodynamic equations to dynamic equations for the
wave function needs a general integration of hydrodynamic equations. This
integration is simple in the partial case of irrotational flow, but it is a
rather complicated mathematical problem in the general case, when a result
of integration has to contain three arbitrary functions of three arguments.
Without producing this integration one cannot derive description of a fluid
in terms of the wave function, and one cannot manipulate dynamic equations,
transforming them from representation in terms of $\rho $, $\mathbf{v}$ to
representation in terms of wave function and back. This problem has not been
solved for years. It had been solved in the end of eighties, and the first
application of this integration can be found in \cite{R89}. Systematical
application of this integration for description of quantum phenomena began
in 1995 \cite{R95,R995}.

The problem of general integration of four hydrodynamic Euler equations 
\begin{eqnarray}
\partial _{0}\rho +\mathbf{\nabla }\left( \rho \mathbf{v}\right) &=&0
\label{a0.2} \\
\partial _{0}\mathbf{v+}\left( \mathbf{v\nabla }\right) \mathbf{v} &=&%
\mathbf{-}\frac{1}{\rho }\mathbf{\nabla }p,\qquad p=p\left( \rho ,\mathbf{%
\nabla }\rho \right)  \label{a0.3}
\end{eqnarray}
seems to be hopeless. It is really so, if the Euler system (\ref{a0.2}), (%
\ref{a0.3}) is considered to be a complete system of dynamic equations. In
fact, the Euler equations (\ref{a0.2}), (\ref{a0.3}) do not form a complete
system of dynamic equations, because it does not describe motion of fluid
particles along their trajectories. To obtain the complete system of dynamic
equations, we should add to the Euler system so called Lin constraints \cite
{L63}

\begin{equation}
\partial _{0}\mathbf{\xi }+\left( \mathbf{v\nabla }\right) \mathbf{\xi }=0
\label{a0.4}
\end{equation}
where $\mathbf{\xi }=\mathbf{\xi }\left( t,\mathbf{x}\right) =\left\{ \xi
_{1},\xi _{2},\xi _{3}\right\} $ are three independent integrals of dynamic
equations 
\[
\frac{d\mathbf{x}}{dt}=\mathbf{v}\left( t,\mathbf{x}\right) , 
\]
describing motion of fluid particles along their trajectories.

Seven equations (\ref{a0.2}) -- (\ref{a0.4}) form the complete system of
dynamic equations, whereas four Euler equations (\ref{a0.2}), (\ref{a0.3})
form only a closed subsystem of the complete system of dynamic equations.
The wave function is expressed via hydrodynamic potentials $\mathbf{\xi }%
=\left\{ \xi _{1},\xi _{2},\xi _{3}\right\} $\textbf{, }which are known also
as Clebsch potentials \cite{C57,C59}. In general case of arbitrary fluid
flow in three-dimensional space the complex wave function $\psi $ has two
complex components $\psi _{1}$, $\psi _{2}$ (or three independent real
components) 
\begin{equation}
\psi =\left( 
\begin{array}{c}
\psi _{1} \\ 
\psi _{2}
\end{array}
\right) =\left( 
\begin{array}{c}
\sqrt{\rho }e^{i\varphi }u_{1}\left( \mathbf{\xi }\right) \\ 
\sqrt{\rho }e^{i\varphi }u_{2}\left( \mathbf{\xi }\right)
\end{array}
\right) ,\qquad \left| u_{1}\right| ^{2}+\left| u_{2}\right| ^{2}=1
\label{a0.5}
\end{equation}

It is impossible to obtain general solution of the Euler system (\ref{a0.2}%
), (\ref{a0.3}), but one can partially integrate the complete system (\ref
{a0.2}) -- (\ref{a0.4}), reducing its order to four dynamic equations for
the wave function (\ref{a0.5}). Practically it means that one integrates
dynamic equations (\ref{a0.4}), where the function $\mathbf{v}\left( t,%
\mathbf{x}\right) $ is determined implicitly by equations (\ref{a0.2}), (\ref
{a0.3}). Such an integration and reduction of the order of the complete
system of dynamic equations appears to be possible, because the system (\ref
{a0.2}) -- (\ref{a0.4}) has the symmetry group, connected with
transformations of the Clebsch potentials 
\begin{equation}
\xi _{\alpha }\rightarrow \tilde{\xi}_{\alpha }=\tilde{\xi}_{\alpha }\left( 
\mathbf{\xi }\right) ,\qquad \alpha =1,2,3,\qquad \frac{\partial \left( 
\tilde{\xi}_{1},\tilde{\xi}_{2},\tilde{\xi}_{3}\right) }{\partial \left( \xi
_{1},\xi _{2},\xi _{3}\right) }\neq 0  \label{a0.6}
\end{equation}
Being a finite function of Clebsch potentials $\mathbf{\xi }$\textbf{, }the
wave function contains information on motion of fluid particles along their
trajectories. This information cannot be obtained from the Euler equations (%
\ref{a0.2}), (\ref{a0.3}) without using Lin constraints (\ref{a0.4}).

Dependence on $\mathbf{\xi }$ takes place only in the case, when the wave
function has more, than one component. In the case of the Schr\"{o}dinger
equation for the spinless particle the wave function has only one component $%
\psi =\sqrt{\rho }e^{i\varphi }u_{1}\left( \mathbf{\xi }\right) $, where in
virtue of the second equation (\ref{a0.5}) $u_{1}\left( \mathbf{\xi }\right)
=1$, and the wave function does not depend on $\mathbf{\xi }$. See details
in \cite{R999,R99}.

After this integration it became clear, that the wave function is only a
method of the fluid description. Application of the wave function as a
fundamental object of quantum mechanics is too restrictive, and its meaning
is obscure, if it is determined axiomatically by quantum principles. The
meaning of the wave function becomes clear, when it is considered to be a
method of the fluid description. Two possible conceptions of the quantum
phenomena are described by the following scheme 
\[
\begin{tabular}{llllll}
$
\begin{array}{c}
\text{axiomatic} \\ 
\text{conception}
\end{array}
:$ & $
\begin{array}{c}
\text{quantum} \\ 
\text{principles}
\end{array}
$ & $\Rightarrow $ & $
\begin{array}{c}
\text{wave } \\ 
\text{function}
\end{array}
$ & $\Rightarrow $ & $
\begin{array}{c}
\text{hydrodynamic} \\ 
\text{interpretation}
\end{array}
$ \\ 
&  &  &  &  &  \\ 
$
\begin{array}{c}
\text{model } \\ 
\text{conception}
\end{array}
:$ & $
\begin{array}{c}
\text{statistical} \\ 
\text{description}
\end{array}
$ & $\Rightarrow $ & $
\begin{array}{c}
\text{fluid} \\ 
\text{dynamics}
\end{array}
$ & $\Rightarrow $ & $
\begin{array}{c}
\text{wave } \\ 
\text{function}
\end{array}
$%
\end{tabular}
\]

Quantum principles are foundation of the conventional quantum theory, and
axiomatic properties of the wave function form a foundation of the
hydrodynamic interpretation, i.e. the wave function is primary and
hydrodynamic interpretation is secondary. In the model description of
quantum phenomena the quantum principles are replaced by the statistical
description, which is connected closely with the fluid dynamics, and fluid
dynamics is a basis for introduction of the wave function, i.e. the fluid
dynamics is primary and the wave function is secondary.

Let $\mathcal{S}_{\mathrm{st}}$ be stochastic particle, whose state $X$ is
described by variables $\left\{ \mathbf{x},\frac{d\mathbf{x}}{dt}\right\} $,
where $\mathbf{x}$ is the particle position. Evolution of the particle state
is stochastic and there exist no dynamic equations for $\mathcal{S}_{\mathrm{%
st}}$. Evolution of the state of $\mathcal{S}_{\mathrm{st}}$ contains both
regular and stochastic components. To separate the regular evolution
components, we consider a set (statistical ensemble) $\mathcal{E}\left[ 
\mathcal{S}_{\mathrm{st}}\right] $ of many independent identical stochastic
particles $\mathcal{S}_{\mathrm{st}}$. All stochastic particles $\mathcal{S}%
_{\mathrm{st}}$ start from the same initial state. It means that all $%
\mathcal{S}_{\mathrm{st}}$ are prepared in the same way. If the number $N$
of $\mathcal{S}_{\mathrm{st}}$ is very large, the stochastic elements of
evolution compensate each other, but regular ones are accumulated. In the
limit $N\rightarrow \infty $ the statistical ensemble $\mathcal{E}\left[ 
\mathcal{S}_{\mathrm{st}}\right] $ turns to a dynamic system, whose state
evolves according to some dynamic equations.

The stochastic particle $\mathcal{S}_{\mathrm{st}}$ has six degrees of
freedom, whereas the statistical ensemble $\mathcal{E}\left[ \mathcal{S}_{%
\mathrm{st}}\right] $ of stochastic particles $\mathcal{S}_{\mathrm{st}}$
has infinite number of freedom degrees. $\mathcal{E}\left[ \mathcal{S}_{%
\mathrm{st}}\right] $ is a continuous dynamic system of the type of a fluid
(continuous medium). If the statistical ensemble $\mathcal{E}\left[ \mathcal{%
S}_{\mathrm{st}}\right] $ is normalized to one particle, it turns to the
statistical average particle $\left\langle \mathcal{S}_{\mathrm{st}%
}\right\rangle $ (See details in \cite{R02}), which is a kind of a fluid.
The action for $\left\langle \mathcal{S}_{\mathrm{st}}\right\rangle $, or
for $\mathcal{E}\left[ \mathcal{S}_{\mathrm{st}}\right] $ can be reduced to
the form of a continuous system $\mathcal{S}_{\mathrm{red}}\left[ \mathcal{S}%
_{\mathrm{d}}\right] $, which consists of many deterministic particles $%
\mathcal{S}_{\mathrm{d}}$, interacting between themselves. Type of this
interaction depends on the character of stochasticity of the particle $%
\mathcal{S}_{\mathrm{st}}$.

Let the statistical ensemble $\mathcal{E}_{\mathrm{d}}\left[ \mathcal{S}_{%
\mathrm{d}}\right] $ of deterministic classical particles $\mathcal{S}_{%
\mathrm{d}}$ be described by the action $\mathcal{A}_{\mathcal{E}_{\mathrm{d}%
}\left[ \mathcal{S}_{\mathrm{d}}\left( P\right) \right] }$, where $P$ are
parameters describing $\mathcal{S}_{\mathrm{d}}$ (for instance, mass,
charge). Let under influence of some stochastic agent the deterministic
particle $\mathcal{S}_{\mathrm{d}}$ turn to a stochastic particle $\mathcal{S%
}_{\mathrm{st}}$. The action $\mathcal{A}_{\mathcal{E}_{\mathrm{st}}\left[ 
\mathcal{S}_{\mathrm{st}}\right] }$ for the statistical ensemble $\mathcal{E}%
_{\mathrm{st}}\left[ \mathcal{S}_{\mathrm{st}}\right] $ is reduced to the
action $\mathcal{A}_{\mathcal{S}_{\mathrm{red}}\left[ \mathcal{S}_{\mathrm{d}%
}\right] }=\mathcal{A}_{\mathcal{E}_{\mathrm{st}}\left[ \mathcal{S}_{\mathrm{%
st}}\right] }$ for some set $\mathcal{S}_{\mathrm{red}}\left[ \mathcal{S}_{%
\mathrm{d}}\right] $ of identical interacting deterministic particles $%
\mathcal{S}_{\mathrm{d}}$. The action $\mathcal{A}_{\mathcal{S}_{\mathrm{red}%
}\left[ \mathcal{S}_{\mathrm{d}}\right] }$ as a functional of $\mathcal{S}_{%
\mathrm{d}}$ has the form $\mathcal{A}_{\mathcal{E}_{\mathrm{d}}\left[ 
\mathcal{S}_{\mathrm{d}}\left( P_{\mathrm{eff}}\right) \right] }$, where
parameters $P_{\mathrm{eff}}$ are parameters $P$ of the deterministic
particle $\mathcal{S}_{\mathrm{d}}$, averaged over the statistical ensemble,
and this averaging describes interaction of particles $\mathcal{S}_{\mathrm{d%
}}$ in the set $\mathcal{S}_{\mathrm{red}}\left[ \mathcal{S}_{\mathrm{d}}%
\right] $. It means that 
\begin{equation}
\mathcal{A}_{\mathcal{E}_{\mathrm{st}}\left[ \mathcal{S}_{\mathrm{st}}\right]
}=\mathcal{A}_{\mathcal{S}_{\mathrm{red}}\left[ \mathcal{S}_{\mathrm{d}%
}\left( P\right) \right] }=\mathcal{A}_{\mathcal{E}_{\mathrm{d}}\left[ 
\mathcal{S}_{\mathrm{d}}\left( P_{\mathrm{eff}}\right) \right] }
\label{a0.6a}
\end{equation}
In other words, stochasticity of particles $\mathcal{S}_{\mathrm{st}}$ in
the ensemble $\mathcal{E}_{\mathrm{st}}\left[ \mathcal{S}_{\mathrm{st}}%
\right] $ is replaced by interaction of $\mathcal{S}_{\mathrm{d}}$ in $%
\mathcal{S}_{\mathrm{red}}\left[ \mathcal{S}_{\mathrm{d}}\right] $, and this
interaction is described by a change 
\begin{equation}
P\rightarrow P_{\mathrm{eff}}  \label{a0.6b}
\end{equation}
in the action $\mathcal{A}_{\mathcal{E}_{\mathrm{d}}\left[ \mathcal{S}_{%
\mathrm{d}}\left( P\right) \right] }$.

How to determine the change (\ref{a0.6b})? At this stage of investigation
the change (\ref{a0.6b}) is phenomenological. In general, the form of the
change (\ref{a0.6b}) is determined by the properties (character) of
stochasticity. But we do not know how to describe the type of stochasticity
and their properties. Then the form of change (\ref{a0.6b}) labels the type
of stochasticity, and the change (\ref{a0.6b}) may be considered to be a
method of the stochasticity description. Before the further discussion let
us consider an example of quantum stochasticity.

Let $\mathcal{S}_{\mathrm{d}}$ be a free deterministic nonrelativistic
particle of the mass $m$. The action for $\mathcal{S}_{\mathrm{d}}$ has the
form 
\begin{equation}
\mathcal{S}_{\mathrm{d}}:\qquad \mathcal{A}\left[ \mathbf{x}\right] =\int
L\left( \mathbf{x,}\frac{d\mathbf{x}}{dt}\right) dt,  \label{a0.7}
\end{equation}
\begin{equation}
L\left( \mathbf{x,}\frac{d\mathbf{x}}{dt}\right) =-mc^{2}+\frac{m}{2}\left( 
\frac{d\mathbf{x}}{dt}\right) ^{2}  \label{a0.8}
\end{equation}
where $\mathbf{x}=\mathbf{x}\left( t\right) $ and $c$ is the speed of the
light.

The action for the ensemble $\mathcal{E}_{\mathrm{d}}\left[ \mathcal{S}_{%
\mathrm{d}}\right] $ of free deterministic particles $\mathcal{S}_{\mathrm{d}%
}$ has the form 
\begin{equation}
\mathcal{E}_{\mathrm{d}}\left[ \mathcal{S}_{\mathrm{d}}\right] :\qquad 
\mathcal{A}_{\mathcal{E}_{\mathrm{d}}\left[ \mathcal{S}_{\mathrm{d}}\left(
m\right) \right] }\left[ \mathbf{x}\right] =\int L\left( \mathbf{x,}\frac{d%
\mathbf{x}}{dt}\right) dtd\mathbf{\xi },  \label{a0.9}
\end{equation}
where the Lagrangian function has the same form (\ref{a0.8}), but $\mathbf{x}%
=\mathbf{x}\left( t,\mathbf{\xi }\right) $, $\mathbf{\xi }=\left\{ \xi
_{1},\xi _{2},\xi _{3}\right\} $. Here variables $\mathbf{\xi }$ label the
particles $\mathcal{S}_{\mathrm{d}}$ of the statistical ensemble $\mathcal{E}%
_{\mathrm{d}}\left[ \mathcal{S}_{\mathrm{d}}\right] $. The action (\ref{a0.9}%
) describes some fluid without pressure.

In the case of a free noncharged particle the only parameter of the particle
is its mass $m$. At the quantum stochasticity the change (\ref{a0.6b}) has
the form 
\begin{equation}
m\rightarrow m_{\mathrm{eff}}=m\left( 1-\frac{\mathbf{u}^{2}}{2c^{2}}+\frac{%
\hbar }{2mc^{2}}\mathbf{\nabla u}\right)  \label{a0.9a}
\end{equation}
where $\mathbf{u}=\mathbf{u}\left( t,\mathbf{x}\right) $ is the mean value
of the stochastic velocity component. Quantum constant $\hbar $ appears here
as a coupling constant, describing connection between the regular and
stochastic components of particle motion. The velocity $\mathbf{u}$ is
supposed to be small with respect to $c$. Then we must make the change (\ref
{a0.9a}) only in the first term of (\ref{a0.8}), because the same change in
the second term give the quantity of the order $O\left( c^{-2}\right) $.
After change (\ref{a0.9a}) the action (\ref{a0.9}) turns to the action

\begin{equation}
\mathcal{E}_{\mathrm{st}}\left[ \mathcal{S}_{\mathrm{st}}\right] :\qquad 
\mathcal{A}_{\mathcal{E}_{\mathrm{d}}\left[ \mathcal{S}_{\mathrm{d}}\left(
m_{\mathrm{eff}}\right) \right] }\left[ \mathbf{x,u}\right] =\int L\left( 
\mathbf{x,}\frac{d\mathbf{x}}{dt}\right) +L_{\mathrm{st}}\left( \mathbf{u},%
\mathbf{\nabla u}\right) dtd\mathbf{\xi },  \label{a0.10}
\end{equation}
\begin{equation}
L_{\mathrm{st}}\left( \mathbf{u},\mathbf{\nabla u}\right) =\frac{m}{2}%
\mathbf{u}^{2}-\frac{\hbar }{2}\mathbf{\nabla u}  \label{a0.11}
\end{equation}
where $\mathbf{x}=\mathbf{x}\left( t,\mathbf{\xi }\right) $, $\mathbf{\xi }%
=\left\{ \xi _{1},\xi _{2},\xi _{3}\right\} $, but $\mathbf{u}=\mathbf{u}%
\left( t,\mathbf{x}\right) $ is a function of $t,\mathbf{x}$. The action (%
\ref{a0.10}) also describes a fluid, but now it is a fluid with a pressure
and its irrotational flow is described by the Schr\"{o}dinger equation \cite
{R02,R002}.

The action (\ref{a0.10}) admits a simple interpretation. The derivative $d%
\mathbf{x/}dt$ describes the mean velocity of a fluid particle which does
not coincide with the velocity of stochastic particle $\mathcal{S}_{\mathrm{%
st}}$, as well as the velocity of a gas particle does not coincide with the
velocity of a gas molecule. The gas molecule has stochastic velocity
component. Likewise, the stochastic particle $\mathcal{S}_{\mathrm{st}}$ has
the stochastic velocity component. The mean value of this stochastic
component is described by the variable $\mathbf{u}$. For a stochastic
Brownian particle the mean value of the stochastic component velocity is
described by the relation 
\begin{equation}
\mathbf{u}_{\mathrm{Br}}=-D\mathbf{\nabla }\ln \rho _{\mathrm{Br}}
\label{a0.12}
\end{equation}
where $D$ is the diffusion coefficient, and $\rho _{\mathrm{Br}}$ is the
density of randomly wandering Brownian particles. Variation of action (\ref
{a0.10}) with respect to $\mathbf{u}$ gives the same result 
\begin{equation}
\mathbf{u=}-\frac{\hbar }{2m}\mathbf{\nabla }\ln \rho ,\qquad \rho =\frac{%
\partial \left( \xi _{1},\xi _{2},\xi _{3}\right) }{\partial \left(
x^{1},x^{2},x^{3}\right) }  \label{a0.13}
\end{equation}
where $\hbar /2m$ may be regarded as the diffusion coefficient.

To obtain the relation (\ref{a0.13}), we need to rewrite (\ref{a0.10}) in
the form of integral over variables $\mathbf{x}$\textbf{, }because $\mathbf{u%
}$ is a function of $\mathbf{x}$ (but not of $\mathbf{\xi })$. 
\begin{equation}
\int L_{\mathrm{st}}\left( \mathbf{u},\mathbf{\nabla u}\right) dtd\mathbf{%
\xi =}\int \left( \frac{m\mathbf{u}^{2}}{2}-\frac{\hbar }{2}\mathbf{\nabla u}%
\right) \rho dtd\mathbf{x},\qquad \rho =\frac{\partial \left( \xi _{1},\xi
_{2},\xi _{3}\right) }{\partial \left( x^{1},x^{2},x^{3}\right) }
\label{a0.14}
\end{equation}
Then variation of (\ref{a0.14}) with respect to $\mathbf{u}$ gives the
relation (\ref{a0.13}). The nature of stochastic velocities $\mathbf{u}$ and 
$\mathbf{u}_{\mathrm{Br}}$ is similar. Both velocities $\mathbf{u}$ and $%
\mathbf{u}_{\mathrm{Br}}$ are a result of random wandering. However, dynamic
equations are different, because dissipative Brownian particle obeys the
Aristotelian dynamics, whereas conservative quantum particle $\mathcal{S}_{%
\mathrm{st}}$ obeys the Newtonian dynamics. The state of the statistical
average Brownian particle $\left\langle \mathcal{S}_{\mathrm{Br}%
}\right\rangle $ is described by the variable $\rho _{\mathrm{Br}}$.
According to (\ref{a0.12}) the velocity $\mathbf{u}_{\mathrm{Br}}$ is a
function of the state $\rho _{\mathrm{Br}}$, and dynamic equation for the
state $\rho _{\mathrm{Br}}$ has the form of the continuity equation 
\[
\frac{d\rho _{\mathrm{Br}}}{\partial t}+\mathbf{\nabla }\left( \rho _{%
\mathrm{Br}}\mathbf{u}_{\mathrm{Br}}\right) =0 
\]

The state of the statistical average particle $\left\langle \mathcal{S}_{%
\mathrm{st}}\right\rangle $ is described by the variables $\rho ,\frac{d%
\mathbf{x}}{dt},\mathbf{u}$, and the energy $m\mathbf{u}^{2}/2$ of the
stochastic component must be taken into account in the Lagrangian function.

The first term $\frac{m}{2}\mathbf{u}^{2}$ of $L_{\mathrm{st}}$ is the
kinetic energy of the stochastic component. But this term does not contain
time derivative. It depends only on $\mathbf{x}$. It means that it acts on
the particle motion as a potential energy. Indeed, variation of (\ref{a0.10}%
) with respect to $\mathbf{x}$ gives dynamic equation 
\[
m\frac{d^{2}\mathbf{x}}{dt^{2}}=-\mathbf{\nabla }U\left( \mathbf{x}\right) 
\mathbf{,\qquad }U\left( \mathbf{x}\right) =\left( \frac{m\mathbf{u}^{2}}{2}-%
\frac{\hbar }{2}\mathbf{\nabla u}\right) 
\]
The second term in (\ref{a0.14}) ensures connection between $\mathbf{u}$ and
collective variable $\rho $.

For application to relativistic particle the relation (\ref{a0.9a}) should
be rewritten in the form

\begin{equation}
m^{2}\rightarrow m_{\mathrm{eff}}^{2}=m^{2}\left( 1-\frac{\mathbf{u}^{2}}{%
c^{2}}+\frac{\hbar }{mc^{2}}\mathbf{\nabla u}\right)  \label{a0.15}
\end{equation}
Besides we should additionally replace nonrelativistic expressions $\mathbf{u%
}^{2}$ and $\mathbf{\nabla u}$ by relativistic ones $-g_{ik}u^{i}u^{k}$ and $%
\partial _{i}u^{i}$ respectively, where $u^{i}$ is the 4-vector, describing
the mean value of the stochastic component of velocity. Then the action (\ref
{a0.8}), (\ref{a0.10}), (\ref{a0.11}) transforms to the relativistic form 
\begin{equation}
\mathcal{A}\left[ x,\kappa \right] =-\int mcK\sqrt{g_{ik}\dot{x}^{i}\dot{x}%
^{k}}d\xi _{0}d\mathbf{\xi ,\qquad }K\mathbf{=}\sqrt{1+\frac{\hbar ^{2}}{%
m^{2}c^{2}}\left( \kappa ^{l}\kappa _{l}+\partial _{l}\kappa ^{l}\right) }
\label{a0.16}
\end{equation}
where $x=\left\{ x^{i}\right\} ,$ $i=0,1,2,3$ is a function of variables $%
\xi =\left\{ \xi _{0},\mathbf{\xi }\right\} =\left\{ \xi _{i}\right\} ,$ $%
\;i=0,1,2,3$ $\dot{x}^{i}\equiv dx^{i}/d\xi _{0}$. Variables $\kappa
=\left\{ \kappa ^{i}\right\} ,$ $\;i=0,1,2,3$ are functions of $x$. The
variables $\kappa ^{l}$ are connected with the variables $u^{l}$ by means of
relation 
\begin{equation}
u^{l}=\frac{\hbar }{m}\kappa ^{l},\qquad l=0,1,2,3  \label{a0.17}
\end{equation}

As well as the nonrelativistic velocity $\mathbf{u}$\textbf{, }the vector
field $\kappa _{l}=g_{lk}\kappa ^{k}$ has a potential $\kappa $ and can be
represented in the form 
\begin{equation}
\kappa _{l}=\partial _{l}\kappa =\frac{1}{2}\partial _{l}\ln \rho ,
\label{a0.18}
\end{equation}
where 
\begin{equation}
\rho =\frac{J\sqrt{\dot{x}_{s}\dot{x}^{s}}}{\rho _{0}mcK}=\frac{\sqrt{%
j_{s}j^{s}}}{\rho _{0}mcK},\qquad j^{k}=J\dot{x}^{k}\equiv \frac{\partial J}{%
\partial \xi _{0,k}}\equiv \frac{\partial \left( x^{k},\xi _{1},\xi _{2},\xi
_{3}\right) }{\partial \left( x^{0},x^{1},x^{2},x^{3}\right) },
\label{a0.19}
\end{equation}
\begin{equation}
J=\frac{\partial \left( \xi _{0},\xi _{1},\xi _{2},\xi _{3}\right) }{%
\partial \left( x^{0},x^{1},x^{2},x^{3}\right) }=\det \left\| \xi
_{i,k}\right\| ,\qquad \xi _{i,k}=\partial _{k}\xi _{i},\qquad i,k=0,1,2,3
\label{a0.20}
\end{equation}
$\rho _{0}=$const.

Result (\ref{a0.18}) is obtained by variation of (\ref{a0.16}) with respect
to $\kappa ^{l}$ in the same way, as equation (\ref{a0.13}) was derived from
(\ref{a0.10}).

The action (\ref{a0.16}) describes some relativistic fluid, whose particles
interact via the self-consistent vector field $\kappa ^{l}$. Properties of
this fluid were investigated in \cite{R98}. It has been shown that
irrotational flow of this fluid is described by the Klein-Gordon equation.
Deterministic relativistic particles $\mathcal{S}_{\mathrm{d}}$ of the set
(fluid) $\mathcal{S}_{\mathrm{red}}\left[ \mathcal{S}_{\mathrm{d}}\right] $
interact between themselves via self-consistent vector field $\kappa ^{l}$, $%
i=0,1,2,3$.

We shall refer to the field $\kappa ^{l}$ and its potential $\kappa $,
defined by the relation (\ref{a0.17}) as $\kappa $-field. We shall show that
the $\kappa $-field has some unusual properties. In particular, the $\kappa $%
-field can produce pairs.

\section{The field producing pairs}

To show that the $\kappa $-field enables to produce pairs, we shall consider
the expression $\frac{\hbar ^{2}}{m^{2}c^{2}}\left( \kappa ^{l}\kappa
_{l}+\partial _{l}\kappa ^{l}\right) $ constructed of the $\kappa $-field as
some given external field $f$. We consider motion of a charged deterministic
particle in the given electromagnetic field $A_{i}$ and some given scalar
field $f$, changing the particle mass. The action for the particle has the
form of the type (\ref{a0.16}) 
\begin{equation}
\mathcal{A}\left[ q\right] =\int L\left( q,\dot{q}\right) d\tau ,\qquad L=-%
\sqrt{m^{2}c^{2}\left( 1+f\left( q\right) \right) g_{ik}\dot{q}^{i}\dot{q}%
^{k}}-\frac{e}{c}A_{k}\dot{q}^{k}  \label{a1.1}
\end{equation}
where relations $x^{i}=q^{i}\left( \tau \right) ,$ $i=0,1,2,3$ describe the
world line of the particle, and $\dot{q}^{k}\equiv dq^{i}/d\tau $. The
quantities $A_{k}=A_{k}\left( q\right) ,$ $k=0,1,2,3$ are given
electromagnetic potentials, and $f=f\left( q\right) $ is some given field,
changing the effective particle mass $m_{\mathrm{eff}}=m\sqrt{\left(
1+f\left( q\right) \right) }$. The canonical momentum $p_{k}$ is defined by
the relation 
\begin{equation}
p_{k}=\frac{\partial L}{\partial \dot{q}^{k}}=-\frac{mcKg_{ki}\dot{q}^{i}}{%
\sqrt{g_{ls}\dot{q}^{l}\dot{q}^{s}}}-\frac{e}{c}A_{k},\qquad K=\sqrt{\left(
1+f\left( q\right) \right) }  \label{a1.2}
\end{equation}
Dynamic equations have the form 
\begin{equation}
\frac{dp_{k}}{d\tau }=-mc\sqrt{g_{ik}\dot{q}^{i}\dot{q}^{k}}\frac{\partial K%
}{\partial q^{k}}-\frac{e}{c}\frac{\partial A_{i}}{\partial q^{k}}\dot{q}^{i}
\label{a1.3}
\end{equation}
The action (\ref{a1.1}), as well as dynamic equations (\ref{a1.3}), (\ref
{a1.2}) are invariant with respect to a transformation of the world line
parametrization 
\begin{equation}
\tau \rightarrow \tilde{\tau}=\tilde{\tau}\left( \tau \right)  \label{a1.4}
\end{equation}
provided $\partial \tilde{\tau}/\partial \tau >0$. If $\partial \tilde{\tau}%
/\partial \tau <0$ the dynamic equations (\ref{a1.3}), (\ref{a1.2}) stop to
be invariant with respect to transformation (\ref{a1.4}) of the world line
parametrization. The parametrization of the world line is a method of the
world line description, and correctly written dynamic equations have not to
depend on the method of description.

Let us modify the action (\ref{a1.1}), introducing orientation $\mathbf{%
\varepsilon }$ of the world line. The orientation of the world line is a
unit vector $\mathbf{\varepsilon }$, tangent to the world line. Component $%
\varepsilon $ of orientation $\mathbf{\varepsilon }$ is the projection of
the vector $\mathbf{\varepsilon }$ on to the vector $dq^{i}/d\tau $%
\begin{equation}
\varepsilon =\frac{g_{ik}\varepsilon ^{i}\dot{q}^{k}}{\sqrt{g_{ls}\dot{q}^{l}%
\dot{q}^{s}}}=\pm 1  \label{a1.4a}
\end{equation}
At the parametrization transformation (\ref{a1.4}) the component $%
\varepsilon $ of orientation $\mathbf{\varepsilon }$ transforms as follows 
\begin{equation}
\varepsilon \rightarrow \tilde{\varepsilon}=\varepsilon \mathrm{sgn}\left(
\partial \tilde{\tau}/\partial \tau \right) ,\qquad \mathrm{sgn}\left(
x\right) =\frac{x}{\left| x\right| }  \label{a1.5}
\end{equation}
Component $\varepsilon $ of orientation $\mathbf{\varepsilon }$ is invariant
with respect to coordinate transformation as it follows from (\ref{a1.4a}).
Let us rewrite the relations (\ref{a1.1}) -- (\ref{a1.3}) in the form, which
is invariant with respect to the arbitrary transformation (\ref{a1.4}) 
\begin{equation}
\mathcal{A}\left[ q\right] =\int\limits_{\min \left( \tau ^{\prime },\tau
^{\prime \prime }\right) }^{\max \left( \tau ^{\prime },\tau ^{\prime \prime
}\right) }L\left( q,\dot{q}\right) d\tau ,\qquad L=-\sqrt{m^{2}c^{2}\left(
1+f\left( q\right) \right) g_{ik}\dot{q}^{i}\dot{q}^{k}}-\frac{\varepsilon e%
}{c}A_{k}\dot{q}^{k}  \label{a1.6}
\end{equation}
where $\tau ^{\prime }$ and $\tau ^{\prime \prime }$ are values of the
parameter $\tau $ at the ends of the integration interval. The particle
momentum is defined 
\begin{equation}
p_{k}=\frac{\partial L}{\partial \dot{q}^{k}}=-\frac{mcKg_{ki}\dot{q}^{i}}{%
\sqrt{g_{ls}\dot{q}^{l}\dot{q}^{s}}}-\frac{\varepsilon e}{c}A_{k},\qquad K=%
\sqrt{\left( 1+f\left( q\right) \right) }  \label{a1.7}
\end{equation}
\begin{equation}
\frac{dp_{k}}{d\tau }=-mc\sqrt{g_{ik}\dot{q}^{i}\dot{q}^{k}}\frac{\partial K%
}{\partial q^{k}}-\frac{\varepsilon e}{c}\frac{\partial A_{i}}{\partial q^{k}%
}\dot{q}^{i}  \label{a1.8}
\end{equation}
Note that now the momentum (\ref{a1.7}) and dynamic equations (\ref{a1.8})
are invariant with respect to arbitrary transformation (\ref{a1.4}). One can
see from (\ref{a1.7}), that the vector 
\begin{equation}
\dot{q}_{k}=\sqrt{\frac{g_{ls}\dot{q}^{l}\dot{q}^{s}}{1+f\left( q\right) }}%
\frac{\left( p_{k}+\frac{\varepsilon e}{c}A_{k}\right) }{mc}  \label{a1.10}
\end{equation}
becomes to be spacelike $\left( g_{ls}\dot{q}^{l}\dot{q}^{s}<0\right) $, if $%
f\left( q\right) <-1$, because only in this case the expression under
radical in (\ref{a1.10}) is real.

The Jacobi-Hamilton equation for the action (\ref{a1.6}) has the form

\begin{equation}
g^{ik}\left( \frac{\partial S}{\partial q^{i}}+\frac{\varepsilon e}{c}%
A_{i}\right) \left( \frac{\partial S}{\partial q^{k}}+\frac{\varepsilon e}{c}%
A_{k}\right) =m^{2}c^{2}\left( 1+f\left( q\right) \right)  \label{a1.9}
\end{equation}

Let us consider solution of the Hamilton-Jacobi equation for the
two-dimensional space-time $\left( t,x\right) $, when $A_{i}=0$, and $%
f=f\left( t\right) $ is a function of only time $t$. In this case the
solution of equation (\ref{a1.9}) has the form 
\begin{equation}
S\left( t,x,p_{0}\right) =p_{0}x+\int\limits_{0}^{t}c\sqrt{m^{2}c^{2}\left(
1+f\left( t\right) \right) +p_{0}^{2}}dt+C\mathbf{,\qquad }p_{0},C=\text{%
const}  \label{a1.11}
\end{equation}
and solution of dynamic equations takes the form 
\begin{equation}
\frac{\partial S\left( t,x,p_{0}\right) }{\partial p_{0}}-x_{0}=x-x_{0}+\int%
\limits_{0}^{t}\frac{p_{0}cdt}{\sqrt{m^{2}c^{2}\left( 1+f\left( t\right)
\right) +p_{0}^{2}}}=0,\qquad x_{0}=\text{const}  \label{a1.12}
\end{equation}

Let for example 
\begin{equation}
f\left( t\right) =\left\{ 
\begin{array}{lll}
0 & \text{if} & t<0 \\ 
-\frac{V^{2}}{m^{2}c^{4}t_{0}^{2}}t\left( t-t_{0}\right) & \text{if} & 
0<t<t_{0} \\ 
0 & \text{if} & t_{0}<t
\end{array}
\right. ,\qquad t_{0},V=\text{const}  \label{a1.13}
\end{equation}
The solution (\ref{a1.12}) takes the form 
\begin{equation}
x=\left\{ 
\begin{array}{lll}
x_{0}-\frac{p_{0}c^{2}}{E}t & \text{if} & t<0 \\ 
x_{0}-\int\limits_{0}^{t}\frac{p_{0}cdt}{\sqrt{E^{2}-V^{2}t\left(
t-t_{0}\right) /t_{0}^{2}}} & \text{if} & 0<t<t_{0} \\ 
x_{1}+\alpha \frac{p_{0}c^{2}}{E}\left( t-t_{0}\right) & \text{if} & t_{0}<t
\end{array}
\right. ,\qquad E=c\sqrt{m^{2}c^{2}+p_{0}^{2}}  \label{a1.14}
\end{equation}
where $\alpha =\pm 1$. Sign of $\alpha $ and the constant $x_{1}$ are
determined from the continuity condition of the world line at $t=t_{0}$. The
solution (\ref{a1.14}) has different form, depending on the sign of the
constant $4E^{2}-V^{2}$.

If $4E^{2}>V^{2}$, the solution (\ref{a1.14}) takes the form 
\begin{equation}
x=\left\{ 
\begin{array}{ll}
x_{0}-\frac{p_{0}c^{2}}{E}t & \text{if }t<0 \\ 
x_{0}-\frac{p_{0}c^{2}t_{0}}{V}\arcsin \frac{2V\left( \sqrt{%
E^{2}t_{0}^{2}-V^{2}t\left( t-t_{0}\right) }-E\left( t_{0}-2t\right) \right) 
}{t_{0}\left( 4E^{2}+V^{2}\right) } & \text{if }0<t<t_{0} \\ 
x_{0}-p_{0}c^{2}\frac{t_{0}}{V}\arcsin \frac{4EV}{4E^{2}+V^{2}}-\frac{%
p_{0}c^{2}}{E}\left( t-t_{0}\right) & \text{if }t_{0}<t
\end{array}
\right. ,\qquad E^{2}>V^{2}/4  \label{a1.15}
\end{equation}

In the case, when $4E^{2}<V^{2}$, the world line is reflected from the
region $\Omega _{\mathrm{fb}}$ of the space-time determined by the condition 
$0<t<t_{0}$, and the coordinate $x$ is not a single-valued function of the
time $t$. In this case we use parametric representation for the solution (%
\ref{a1.14}). We have 
\begin{equation}
x=\left\{ 
\begin{array}{lll}
x_{0}-\frac{p_{0}c^{2}t_{0}}{2E}\left( 1-A\cosh \tau \right) & \text{if} & 
\tau <-\tau _{0} \\ 
x_{0}-\frac{p_{0}c^{2}t_{0}}{V}\left( \tau +\tau _{0}\right) & \text{if} & 
-\tau _{0}<\tau <\tau _{0} \\ 
x_{0}-\frac{2p_{0}t_{0}}{V}\tau _{0}+\frac{p_{0}c^{2}t_{0}}{2E}\left(
1-A\cosh \tau \right) & \text{if} & \tau _{0}<\tau
\end{array}
\right.  \label{a1.17}
\end{equation}
\begin{equation}
t=\frac{t_{0}}{2}\left( 1-A\cosh \tau \right)  \label{a1.18}
\end{equation}
where 
\begin{equation}
A=\sqrt{1-\frac{4E^{2}}{V^{2}}},\qquad \tau _{0}=\func{arccosh}\frac{1}{A}=%
\func{arccosh}\frac{1}{\sqrt{1-\frac{4E^{2}}{V^{2}}}}  \label{a1.19}
\end{equation}

The solution (\ref{a1.17}), (\ref{a1.18}) describes annihilation of particle
and antiparticle with the energy $E<V/2$ in the region $0<t<t_{0}$.
Solution, describing the particle-antiparticle generation, has the form 
\begin{equation}
x=\left\{ 
\begin{array}{lll}
x_{0}-\frac{p_{0}c^{2}t_{0}}{2E}\left( A\cosh \tau -1\right) & \text{if} & 
\tau <-\tau _{0} \\ 
x_{0}+\frac{p_{0}c^{2}t_{0}}{V}\left( \tau +\tau _{0}\right) & \text{if} & 
-\tau _{0}<\tau <\tau _{0} \\ 
x_{0}+\frac{2p_{0}t_{0}}{V}\tau _{0}+\frac{p_{0}c^{2}t_{0}}{2E}\left( A\cosh
\tau -1\right) & \text{if} & \tau _{0}<\tau
\end{array}
\right.  \label{a1.20}
\end{equation}
\begin{equation}
t=\frac{t_{0}}{2}\left( A\cosh \tau -1\right)  \label{a1.21}
\end{equation}
where parameters $A,\tau _{0}$ are defined by the relation (\ref{a1.19}),
and the relation $2E<V$ takes place.

In both cases (\ref{a1.17}) and (\ref{a1.20}) at $\left| t\right|
\rightarrow \infty $ the world line has two branches, which can be
approximated by the relations 
\begin{equation}
x=x_{0}+vt_{1}\pm v\left( t-t_{1}\right) ,\qquad t_{1}=t_{0}\frac{E}{V}
\label{a1.22}
\end{equation}
where $v=-\frac{p_{0}c^{2}}{E}$ is the particle velocity, and $v=\frac{%
p_{0}c^{2}}{E}$ is the antiparticle velocity.

Thus, the region $\Omega _{\mathrm{fb}}$ with the $\kappa $-field reflects
world lines of particles with the energy $E<V/2$. If the particle energy $%
E>V/2$, the particle passes the region $\Omega _{\mathrm{fb}}$ through. The
world line penetrates into forbidden region $\Omega _{\mathrm{fb}}$ the
deeper the larger is its energy.

\section{Connection between energy-momentum and \newline
canonical momentum}

It is common practice to think that in the absence of electromagnetic field
the canonical momentum (\ref{a1.7}) coincides with the particle
energy-momentum $P_{i}$. Then the particle energy may be negative, as far as
the canonical momentum component $p_{0}$ may be positive and negative.
Existence of negative particle energy is very undesirable, because it leads
to many very strange corollaries.

In reality, the energy-momentum \ $P_{i}$, $\;i=0,1,2,3$ and canonical
momentum $p_{i}$, $\;i=0,1,2,3$ are different quantities which can coincide
only in the case, when there are only particles, or only antiparticles. They
are defined differently. The canonical momentum $p_{i}$ is defined as the
quantity canonically conjugate to coordinate $x^{i}$ of the particle
position by means of the relation (\ref{a1.7}). The energy-momentum $P_{i}$
is defined by the relation 
\begin{equation}
P_{i}=\int\limits_{V}T_{i}^{0}d\mathbf{x,\qquad }i=0,1,2,3  \label{a2.1}
\end{equation}
where $T_{i}^{0}=g_{ik}T^{k0}$ are components of the energy-momentum tensor,
and $V$ is the 3-volume around the particle, whose energy-momentum is
determined. The energy-momentum tensor $T^{ik}$ is defined as a source of
the gravitational field, i.e. by means of the relation 
\begin{equation}
T^{ik}\left( x\right) =\frac{\delta \mathcal{A}}{\delta g_{ik\left( x\right)
}}=-\frac{2}{\sqrt{-g}}\frac{\partial }{\partial g_{ik}\left( x\right) }%
\left( \sqrt{-g}\mathcal{L}\right)  \label{a2.2}
\end{equation}
where $\mathcal{A}$ is the action (\ref{a1.6}), written in the arbitrary
curvilinear coordinate system in the form 
\begin{equation}
\mathcal{A}\left[ q^{i},g_{ik},A_{i}\right] =\int \mathcal{L}\left(
q^{i},g_{ik},A_{i}\right) \sqrt{-g}d^{4}x  \label{a2.3}
\end{equation}
and $g=\det \left| \left| g_{ik}\right| \right| .$

In the same way the electric charge of a particle is defined by the relation 
\begin{equation}
Q=\int\limits_{V}j_{\mathrm{c}}^{0}d\mathbf{x,}  \label{a2.4}
\end{equation}
where $j_{\mathrm{c}}^{i}$, $i=0,1,2,3$ is electric current density, defined
as the source of the electromagnetic field, i.e. by the relation 
\begin{equation}
j_{\mathrm{c}}^{i}\left( x\right) =c\frac{\delta \mathcal{A}}{\delta
A_{i}\left( x\right) }=\frac{c}{\sqrt{-g}}\frac{\partial }{\partial
A_{i}\left( x\right) }\left( \sqrt{-g}\mathcal{L}\right)  \label{a2.5}
\end{equation}

To obtain expressions for the quantities $j_{\mathrm{c}}^{l}$, $T^{ik}$, let
us write the action (\ref{a1.6}) in the arbitrary curvilinear coordinate
system and represent it in the form (\ref{a2.3}). One obtains

\begin{equation}
\mathcal{A}=\int d^{4}x\int\limits_{\min \left( \tau ^{\prime },\tau
^{\prime \prime }\right) }^{\max \left( \tau ^{\prime },\tau ^{\prime \prime
}\right) }L\left( q,\dot{q}\right) \delta ^{\left( 4\right) }\left(
x-q\left( \tau \right) \right) d\tau ,  \label{a2.6}
\end{equation}
where 
\begin{equation}
\delta ^{\left( 4\right) }\left( x-q\left( \tau \right) \right) =\delta
\left( x^{0}-q^{0}\left( \tau \right) \right) \delta \left( \mathbf{x}-%
\mathbf{q}\left( \tau \right) \right) =\prod_{i=0}^{i=3}\delta \left(
x^{i}-q^{i}\left( \tau \right) \right)  \label{a2.7}
\end{equation}

Integrating (\ref{a2.6}) over $\tau $ and using $\delta $-function $\delta
\left( x^{0}-q^{0}\left( \tau \right) \right) $, one obtains 
\begin{equation}
\mathcal{A}=\int d^{4}x\sum_{j}\frac{L\left( q\left( \tau _{j}\right) ,\dot{q%
}\left( \tau _{j}\right) \right) }{\left| \dot{q}^{0}\left( \tau _{j}\right)
\right| }\delta \left( \mathbf{x}-\mathbf{q}\left( \tau _{j}\right) \right) ,
\label{a2.8}
\end{equation}
where $\tau _{j}=\tau _{j}\left( x^{0}\right) $ are roots of the equation 
\begin{equation}
x^{0}-q^{0}\left( \tau _{j}\right) =0  \label{a2.9}
\end{equation}

Comparing relations (\ref{a2.8}) and (\ref{a2.3}), we conclude 
\begin{eqnarray}
\mathcal{L}\sqrt{-g} &=&\sum_{j}\frac{L\left( q\left( \tau _{j}\right) ,\dot{%
q}\left( \tau _{j}\right) \right) }{\left| \dot{q}^{0}\left( \tau
_{j}\right) \right| }\delta \left( \mathbf{x}-\mathbf{q}\left( \tau
_{j}\right) \right)  \nonumber \\
&=&\sum_{j}\left( -\sqrt{m^{2}c^{2}\left( 1+f\left( q\left( \tau _{j}\right)
\right) \right) g_{ik}\dot{q}^{i}\left( \tau _{j}\right) \dot{q}^{k}\left(
\tau _{j}\right) }-\frac{\varepsilon e}{c}A_{k}\dot{q}^{k}\left( \tau
_{j}\right) \right)  \nonumber \\
&&\times \frac{\delta \left( \mathbf{x}-\mathbf{q}\left( \tau _{j}\right)
\right) }{\left| \dot{q}^{0}\left( \tau _{j}\right) \right| }  \label{a2.10}
\end{eqnarray}
Now let us substitute relation (\ref{a2.10}) in the relations (\ref{a2.2}), (%
\ref{a2.5}) and set\ \ $A_{i}=0$, $\;g_{ik}=$diag$\left\{
c^{2},-1,-1,-1\right\} $. After calculations we obtain relations for the
energy-momentum tensor and the current density 
\begin{equation}
T^{ik}\left( x^{0},\mathbf{x}\right) =\sum_{j}\frac{mc\sqrt{1+f\left(
q\left( \tau _{j}\right) \right) }\dot{q}^{i}\left( \tau _{j}\right) \dot{q}%
^{k}\left( \tau _{j}\right) }{c\sqrt{g_{ls}\dot{q}^{l}\left( \tau
_{j}\right) \dot{q}^{s}\left( \tau _{j}\right) }}\frac{\delta \left( \mathbf{%
x}-\mathbf{q}\left( \tau _{j}\right) \right) }{\left| \dot{q}^{0}\left( \tau
_{j}\right) \right| }  \label{a2.11}
\end{equation}

\begin{equation}
j^{k}_{\mathrm{c}}=-\sum_{j}\varepsilon e\dot{q}^{k}\left( \tau _{j}\right) 
\frac{\delta \left( \mathbf{x}-\mathbf{q}\left( \tau _{j}\right) \right) }{%
\left| \dot{q}^{0}\left( \tau _{j}\right) \right| }  \label{a2.12}
\end{equation}
where $\tau _{j}=\tau _{j}\left( x^{0}\right) $. Substituting relations (\ref
{a2.11}), (\ref{a2.12}) in relations (\ref{a2.1}), (\ref{a2.4}), we obtain 
\begin{eqnarray}
P_{i} &=&\sum_{j}\frac{mc\sqrt{1+f\left( q\left( \tau _{j}\right) \right) }%
\dot{q}_{i}\left( \tau _{j}\right) \dot{q}^{0}\left( \tau _{j}\right) }{%
\left| \dot{q}^{0}\left( \tau _{j}\right) \right| \sqrt{g_{ls}\dot{q}%
^{l}\left( \tau _{j}\right) \dot{q}^{s}\left( \tau _{j}\right) }}  \nonumber
\\
&=&-\sum_{j}\frac{p_{i}\left( \tau _{j}\right) \dot{q}^{0}\left( \tau
_{j}\right) }{\left| \dot{q}^{0}\left( \tau _{j}\right) \right| }%
=-\sum_{j}p_{i}\left( \tau _{j}\right) \mathrm{sgn}\left( \dot{q}^{0}\left(
\tau _{j}\right) \right)  \nonumber \\
&=&\sum_{j}p_{i}\left( \tau _{j}\right) \mathrm{sgn}\left( p_{0}\left( \tau
_{j}\right) \right) ,\qquad i=0,1,2,3  \label{a2.13}
\end{eqnarray}
\begin{equation}
Q=-\sum_{j}\varepsilon e\cdot \mathrm{sgn}\left( \dot{q}^{0}\left( \tau
_{j}\right) \right) =\sum_{j}\varepsilon e\cdot \mathrm{sgn}\left(
p_{0}\left( \tau _{j}\right) \right)  \label{a2.14}
\end{equation}
In particular, we obtain for the energy $E=P_{0}$%
\begin{equation}
P_{0}=E=\sum_{j}p_{0}\left( \tau _{j}\right) \mathrm{sgn}\left( p_{0}\left(
\tau _{j}\right) \right) =\sum_{j}\left| p_{0}\left( \tau _{j}\right) \right|
\label{a2.15}
\end{equation}

In the case, when $f\left( q\right) +1<0$ in some regions of the space-time,
the equation (\ref{a2.9}) may have several roots. Each of roots corresponds
to some intersection of the world line with the plane $t=$const. Any of such
intersections describes either particle, when sgn$\left( \dot{q}^{0}\right)
=-$sgn$\left( p_{0}\right) >0$, or antiparticle, when sgn$\left( \dot{q}%
^{0}\right) =-$sgn$\left( p_{0}\right) <0.$\ According to (\ref{a2.14}) the
electric charge $Q_{\mathrm{p}}$ of a particle is equal to $-\varepsilon e$,
whereas the electric charge $Q_{\mathrm{a}}$ of an antiparticle is equal to $%
\varepsilon e$. Thus, the particle and antiparticle have electric charges of
different signs, because the sign of $p_{0}$ is different for particle and
antiparticle. According to (\ref{a2.15}) the energy is positive for a
particle as well as for an antiparticle, although the sign of the temporal
component $p_{0}$ of the canonical momentum is different for particle and
antiparticle.

Such a situation seems to be very reasonable. It can be interpreted as
follows. Let the term ''SWL'' (section of world line) mean the collective
concept with respect to concepts ''particle'' and ''antiparticle''. The
particle and antiparticle distinguish in the sign of the temporal component $%
p_{0}$ of the canonical momentum and in the sign of the temporal component $%
-K\dot{q}^{0}\left( g_{ls}\dot{q}^{l}\dot{q}^{s}\right) ^{-1/2}$ of the
velocity 4-vector. Flux of the world lines through some spacelike
three-dimensional area $dS_{i}$ for particles and the flux of the world
lines through the same area $dS_{i}$ for antiparticles have different signs.
This fact may be interpreted in the sense that a particle is one SWL,
whereas an antiparticle is minus one SWL. If SWL has the canonical momentum $%
p_{i}$, then the number $n$ of SWLs is $n=-$sgn$\left( p_{0}\right) $. The
relations (\ref{a2.13}) -- (\ref{a2.15}) can be written in the form 
\begin{eqnarray}
Q &=&-\sum_{j}\varepsilon en\left( \tau _{j}\right) ,\qquad
E=-\sum_{j}p_{0}\left( \tau _{j}\right) n\left( \tau _{j}\right)
\label{a2.16} \\
P_{i} &=&-\sum_{j}p_{i}\left( \tau _{j}\right) n\left( \tau _{j}\right)
,\qquad i=0,1,2,3,\qquad n=-\mathrm{sgn}\left( p_{0}\right)  \label{a2.17}
\end{eqnarray}

This is reasonable formulas for additive quantities, when, the additive
quantity $a$ of several objects (SWLs) is a sum of additive quantities $a$
of each object (SWL) multiplied by the number $n$ of objects (SWLs). It is
unusual only that the number $n$ of objects (SWLs) may be negative.

\section{Peculiarities of description in terms of world lines}

From viewpoint of the consistent relativity theory a world line is a real
physical object, whereas a particle and an antiparticle (SWLs) are
attributes of the world line, which appear as a result of intersection of
the world line with the plane $t=$const. We shall use a special term `WL'
for the world line, considered to be a real physical object. In the
nonrelativistic theory a real physical object is a particle (or
antiparticle), whereas the world line is an attribute of a particle (its
history).

Particle and antiparticle in the nonrelativistic theory (or in inconsistent
relativistic theory) are two different objects, having different parameters.
The particle is described by the following parameters $\left\{ m,e\right\} $%
, where $m$ is a mass, and $e$ is a charge constant. The antiparticle is
described by the parameters $\left\{ m,-e\right\} $. Particle and
antiparticle are described by different dynamic systems, because these
systems have different parameters. Note that at such a description the
component $\varepsilon $ of orientation $\mathbf{\varepsilon }$ is not a
parameter of dynamic system. It is a method of description of SWL, because $%
\varepsilon $ changes at the change $\tau \rightarrow -\tau $ of the world
line parametrization. At nonrelativistic approach one cannot describe pair
production classically, because one cannot connect fragments of world lines
of two different objects (particle and antiparticle). In the framework of
classical theory consideration of a particle and an antiparticle as two
different objects is incompatible. In the quantum theory, where the particle
world line may not exist, it is possible to introduce special operators,
describing production and annihilation of a particle (antiparticle). In
quantum theory pair generation is a corollary of dynamic equations, and one
can imagine such a situation (at proper choice of dynamic equations), when
particles and antiparticles are generated separately. In the classical
theory the pair generation of particles and antiparticles is described on
the deeper (descriptive) level. Choice of dynamic equations cannot change
anything.

Nevertheless, let us start from the consistent relativistic theory, where
SWLs are attributes of WL, and imagine that, describing motion of SWLs
classically, we divide the world line $\mathcal{L}$ into fragments $\mathcal{%
L}_{i} $,\ $i=1,2...$in such a way, that each fragment $\mathcal{L}_{i}$
describe either a particle, or an antiparticle. To make this, it is
sufficient to make the following transformation for fragments $\mathcal{L}%
_{i}$, describing antiparticle. 
\begin{equation}
e\rightarrow -e,\qquad \tau \rightarrow -\tau ,\qquad \varepsilon
\rightarrow \varepsilon  \label{a3.1}
\end{equation}

According to (\ref{a2.13}), (\ref{a2.14}) this transformation changes
neither energy-momentum vector $P_{i}$, nor the charge $Q$ of SWL, but
transformation (\ref{a3.1}) changes the sign of the canonical momentum for
antiparticles and does not change it for particles. As a result the temporal
component $p_{0}$ of particles and antiparticles becomes negative, and one
can identify the energy $E$ of particles and antiparticles with the time
component $-p_{0}$ of their canonical momentum, i.e. $E=-p_{0}$.
Conventionally such an identification is always produced, to satisfy so
called causality principle. Temporal component $-p_{0}$ of canonical
momentum is considered to be an energy of both particle and antiparticle.
Using such an identification, it is always supposed that $dt/d\tau >0$, and
one may ignore concept of orientation.

The action (\ref{a1.6}) and dynamic equations (\ref{a1.8}) are invariant
with respect to transformation (\ref{a3.1}), whereas the canonical momentum (%
\ref{a1.7}) changes its sign. As a result all fragments $\mathcal{L}_{i}$ of 
$\mathcal{L}$ have the same sgn$\left( p_{0}\right) $, and $E=-p_{0}$ along
the whole $\mathcal{L}$. At the same time \thinspace different fragments $%
\mathcal{L}_{i}$ of $\mathcal{L}$ (particles and antiparticles) are
described as different dynamic systems with parameters $\left\{ m,e\right\} $
and $\left\{ m,-e\right\} $ respectively.

Then interval between any two points $P$ and $P^{\prime }$, belonging to the
same fragment $\mathcal{L}_{i}$, is timelike (we neglect the short spacelike
segments of $\mathcal{L}_{i}$ near its ends). If interval between two points 
$P_{1}$ and $P_{2}$ is spacelike, then $P_{1}\in \mathcal{L}_{i}$ and $%
P_{2}\in \mathcal{L}_{k}$, $\;i\neq k$. The points belong to different
dynamic systems and the Poisson brackets between dynamic variables at the
points $P_{1}$ and $P_{2}$ vanish 
\begin{equation}
\left\{ u\left( P_{1}\right) ,v\left( P_{2}\right) \right\} =0,\qquad \left(
x_{\left( \mathrm{1}\right) }^{i}-x_{\left( \mathrm{2}\right) }^{i}\right)
\left( x_{\left( \mathrm{1}\right) i}-x_{\left( \mathrm{2}\right) i}\right)
<0  \label{a2.18}
\end{equation}
where $u\left( P_{1}\right) $ and $v\left( P_{2}\right) $ are any dynamic
variables at the point $P_{1}$ and at the point $P_{2}$ respectively. The
quantities $x_{\left( \mathrm{1}\right) }^{i}$ and $x_{\left( \mathrm{2}%
\right) }^{i}$ are coordinates of points $P_{1}$ and $P_{2}$ respectively.

In the quantum field theory, where fragmentation of the total world line $%
\mathcal{L}$ is used, the relation (\ref{a2.18}) takes the form 
\begin{equation}
\left[ u\left( x_{1}\right) ,v\left( x_{2}\right) \right] _{-}=0,\qquad
\left( x_{1}-x_{2}\right) ^{2}<0  \label{a2.19}
\end{equation}
where $\left[ ...\right] _{-}$ means a commutator of two operators,
describing dynamic variables $u$ and $v$ at points $x_{1}$ and $x_{2}$
respectively.

The relation (\ref{a2.19}) is known as the causality principle.
Interpretation of this principle is such. Events at the points $P_{1}$ and $%
P_{2}$ are independent, because no interaction can propagate faster, than
the speed of the light.

If the world line $\mathcal{L}$ is considered to be a whole object (WL),
described by dynamic equations (\ref{a1.7}), (\ref{a1.8}) and parameter $%
\tau $, changing monotonically along $\mathcal{L}$, there are points $P_{1}$
and $P_{2}$, separated by spacelike interval and belonging to $\mathcal{L}$.
Of course, it is possible only, if there are pair production and pair
annihilation. For instance, in the example of the second section (see (\ref
{a1.22})) the points with coordinates $P_{1}=\left( t,x\right) $ and $%
P_{2}=\left( t,2\left( x_{0}+vt_{1}\right) -x\right) $ are separated by
spacelike interval $2i\left| x-x_{0}-vt_{1}\right| $. According to (\ref
{a1.22}) they both belong to the same world line. As far as the state $%
\left( x_{2},p_{2}\right) $ of dynamic system at the point $P_{2}$ is a
function of its state $\left( x_{1},p_{1}\right) $ at the point $P_{1}$, the
relation (\ref{a2.18}) cannot be fulfilled for all functions $u$ and $v$ of
dynamic variables. It means that the causality principle in the form (\ref
{a2.18}) is violated. Events at points $P_{1}$ and $P_{2}$ appear to be
dependent, when the particle and antiparticle are described as different
states of total WL (but not as separate dynamic systems).

Does it mean that the description in terms of WL is incompatible with the
causality principle? No, because description in terms of one WL is only a
part of the complete description of the particle-antiparticle collision.
Indeed, let us imagine that we prepare high energy particle and antiparticle
at the points $P_{1}$ and $P_{2}$, separated by spacelike interval. Let
these SWLs move to meet each other. Then there are the following
possibilities: (1) the particle and antiparticle annihilate (description in
terms of one WL), (2) particle and antiparticle do not collide (description
in terms of two WLs), (3) particle and antiparticle collide and generates $n$
pairs additionally (description in terms of $n+1$ WLs). Thus, the complete
description of particle-antiparticle collision is described statistically by
means of a sum of $k$-WL descriptions $k=1,2,...$ The causality principle
can be applied only to the complete description.

Description in terms of one WL is a special part of the complete
description, and this partial description may contain a correlation between
the points, separated by a spacelike interval, because the fact that the
particle and antiparticle annihilate supposes some correlation between the
states at the points $P_{1}$ and $P_{2}$, separated by a spacelike interval.
This correlation for one-WL description means that not all Poison brackets (%
\ref{a2.18}) do vanish. This correlation does not mean a violation of the
causality principle, because one-WL description does not pretend to a
complete description of the particle-antiparticle collision. The
conventional description in terms of particles and antiparticles pretend to
a complete description, and fulfillment of the causality principle is
essential for the complete description.

Understanding that the complete description is a very complicated problem,
we divide the problem into parts (one WL, two WLs,...) and try to solve any
part separately. Our strategy of investigation reminds the investigation of
interaction of nonrelativistic particles, where the number of particles
conserves. Then the whole investigation can be separated into one-particle
problem, two-particle problem, three-particle problem,...These problems are
solved in series. In the relativistic case the number of WLs is conserved,
and the whole problem can be separated into parts, corresponding to fixed
number of WLs. Some ideas of such an approach one can find in \cite{R94},
but details need further development. In any case a description of the pair
production in terms of one WL does not contradict to the causality
principle, because this description does not pretend to a complete
description of the collision process.

In the modern relativistic quantum field theory a description is produced in
terms of particles and antiparticles (but not in terms of WLs). This
statement follows from identification of the energy operator $E$ with the
Hamilton operator $\ H=p_{0}$ which is defined as an evolution operator.
Formally this identification follows from relations 
\begin{equation}
i\hbar \partial _{k}u=\left[ u,P_{k}\right] _{-},\qquad P_{k}=\int T_{k}^{0}d%
\mathbf{x},\qquad k=0,1,2,3  \label{a3.6}
\end{equation}
where $u$ is an operator of arbitrary dynamic variable, $\left[ ...\right]
_{-}$ denotes commutator, and $P_{k}$ is the energy-momentum operator of the
dynamic system. At such a description the causality principle is valid,
because description is carried out in terms of particles and antiparticles
(but not in terms of WLs).

A difficulty of such an approach is a consideration of indefinite number of
nonconserved objects (particles and antiparticles). This leads a
perturbative description of the pair production processes.

Conceptual defect of conventional description with a use of identification (%
\ref{a3.6}) is incompatibility of the relations (\ref{a3.6}) with the
dynamic equations. Restrictions (\ref{a3.6}) are compatible with dynamic
equations, obtained at the canonical secondary quantization of linear
equation 
\begin{equation}
\hbar ^{2}\partial _{i}\partial ^{i}\psi +m^{2}c^{2}\psi =0  \label{a3.8}
\end{equation}
But they stop to be compatible with dynamic equations, obtained at the
canonical secondary quantization of nonlinear equation 
\begin{equation}
\hbar ^{2}\partial _{i}\partial ^{i}\psi +m^{2}c^{2}\psi =g\psi ^{\ast }\psi
\psi ,\qquad g=\text{const}  \label{a3.9}
\end{equation}

The fact is that the secondary quantization of (\ref{a3.9}) can be produced
without imposition of constraints (\ref{a3.6}). Such a quantization is
produced in terms of WLs \cite{R001}. It does not use the perturbation
theory and admits to obtain solution of the scattering problem without cut
off at $t\rightarrow \pm \infty $. If the secondary quantization can be
carried out without imposition of (\ref{a3.6}), one should test whether the
relations (\ref{a3.6}) are compatible with dynamic equations. The test shows
that the constraints (\ref{a3.6}) are compatible with the dynamic equations
only for linear equation (\ref{a3.8}), but they are incompatible with (\ref
{a3.9}) \cite{R001}. After imposition of constraints (\ref{a3.6}) the
problem becomes overdetermined. The overdetermined problem is inconsistent,
and one can obtain practically any desired result for such a statement of
the problem.

It is quite reasonable that the causality principle (\ref{a2.19}) is not
fulfilled at the quantization in terms of WLs, because it is a corollary of
the constraints (\ref{a3.6}). Secondary quantization in terms WLs \cite{R001}
does not describe pair production. It does not describe even scattering of a
particle on antiparticle. But at the same time this quantization is
consistent and well defined from the formal viewpoint. What does it mean? It
means only that the nonlinear self-action term in (\ref{a3.9}) does not
describe pair production, because the pair production is apparently a more
subtle effect. But why conventionally at quantization \cite
{GJ68,GJ70,GJ970,GJ72} of (\ref{a3.9}) one obtains pair production? The
answer is simple. Because of overdetermination of the problem, that makes
the problem inconsistent. Practically, it looks as follows. Imposition of
the constraint (\ref{a3.9}) leads to fragmentation of the total world line
into particles and antiparticles. After scattering one needs to produce
defragmentation and collect fragments into the total world lines. Process of
defragmentation is imperfect, because of inconsistent perturbative
description. Reminders of imperfect defragmentation imitate pair production.
At the conventional approach any nonlinear term leads to pair production,
whereas classical description of the second section shows that the field,
producing pairs must have rather specific properties.

Failure of the paper \cite{R001} in description of pair production and
inconsistency of conventional quantization in terms of particles and
antiparticles show that one needs to investigate properties of the $\kappa $%
-field, producing pairs, more carefully, trying to understand what factor is
responsible for pair production.

\section{Transformation of dynamic equations}

Let us add to the action (\ref{a0.16}) the term describing interaction with
the electromagnetic field and write it in the form 
\begin{eqnarray}
\mathcal{A}\left[ x,\kappa \right] &=&\int \left\{ -mcK\sqrt{g_{ik}\dot{x}%
^{i}\dot{x}^{k}}-\frac{e}{c}A_{k}\dot{x}^{k}\right\} d^{4}\xi ,\qquad
d^{4}\xi =d\xi _{0}d\mathbf{\xi ,}  \label{A.1} \\
K &=&\sqrt{1+\lambda ^{2}\left( \kappa _{l}\kappa ^{l}+\partial _{l}\kappa
^{l}\right) },\qquad \lambda =\frac{\hbar }{mc}  \label{A.2}
\end{eqnarray}
where $x=\left\{ x^{i}\left( \xi _{0},\mathbf{\xi }\right) \right\} ,\;$ $%
i=0,1,2,3$ are dependent variables. $\xi =\left\{ \xi _{0},\mathbf{\xi }%
\right\} =\left\{ \xi _{k}\right\} ,\;\;k=0,1,2,3$ are independent
variables, and $\dot{x}^{i}\equiv dx^{i}/d\xi _{0}.$ The quantities $\kappa
^{l}=\left\{ \kappa ^{l}\left( x\right) \right\} ,\;$ $l=0,1,2,3$ are
dependent variables, describing stochastic component of the particle motion, 
$A_{k}=\left\{ A_{k}\left( x\right) \right\} ,\;\;k=0,1,2,3$ is the
potential of electromagnetic field. We shall refer to the dynamic system,
described by the action (\ref{A.1}), (\ref{A.2}) as $\mathcal{S}_{\mathrm{KG}%
}$, because irrotational flow of $\mathcal{S}_{\mathrm{KG}}$ is described by
the Klein-Gordon equation \cite{R98}. We present here this transformation to
the Klein-Gordon form, because we shall use intermediate stages of this
transformation for further investigations.

Let us consider variables $\xi =\xi \left( x\right) $ in (\ref{A.1}) as
dependent variables and variables $x$ as independent variables. Let the
Jacobian 
\begin{equation}
J=\frac{\partial \left( \xi _{0},\xi _{1},\xi _{2},\xi _{3}\right) }{%
\partial \left( x^{0},x^{1},x^{2},x^{3}\right) }=\det \left| \left| \xi
_{i,k}\right| \right| ,\qquad \xi _{i,k}\equiv \partial _{k}\xi _{i},\qquad
i,k=0,1,2,3  \label{A.3}
\end{equation}
be considered to be a multilinear function of $\xi _{i,k}$. Then 
\begin{equation}
d^{4}\xi =Jd^{4}x,\qquad \dot{x}^{i}\equiv \frac{dx^{i}}{d\xi _{0}}\equiv 
\frac{\partial \left( x^{i},\xi _{1},\xi _{2},\xi _{3}\right) }{\partial
\left( \xi _{0},\xi _{1},\xi _{2},\xi _{3}\right) }=J^{-1}\frac{\partial J}{%
\partial \xi _{0,i}}  \label{A.4}
\end{equation}
After transformation to dependent variables $\xi $ the action (\ref{A.1})
takes the form 
\begin{equation}
\mathcal{A}\left[ \xi ,\kappa \right] =\int \left\{ -mcK\sqrt{g_{ik}\frac{%
\partial J}{\partial \xi _{0,i}}\frac{\partial J}{\partial \xi _{0,k}}}-%
\frac{e}{c}A_{k}\frac{\partial J}{\partial \xi _{0,k}}\right\} d^{4}x\mathbf{%
,}  \label{A.5}
\end{equation}

Let us introduce new variables 
\begin{equation}
j^{k}=\frac{\partial J}{\partial \xi _{0,k}},\qquad k=0,1,2,3  \label{A.6}
\end{equation}
by means of Lagrange multipliers $p_{k}$%
\begin{equation}
\mathcal{A}\left[ \xi ,\kappa ,j,p\right] =\int \left\{ -mcK\sqrt{%
g_{ik}j^{i}j^{k}}-\frac{e}{c}A_{k}j^{k}+p_{k}\left( \frac{\partial J}{%
\partial \xi _{0,k}}-j^{k}\right) \right\} d^{4}x\mathbf{,}  \label{A.7}
\end{equation}
Variation with respect to $\xi _{i}$ gives 
\begin{equation}
\frac{\delta \mathcal{A}}{\delta \xi _{i}}=-\partial _{l}\left( p_{k}\frac{%
\partial ^{2}J}{\partial \xi _{0,k}\partial \xi _{i,l}}\right) =0,\qquad
i=0,1,2,3  \label{A.8}
\end{equation}
Using identities 
\begin{equation}
\frac{\partial ^{2}J}{\partial \xi _{0,k}\partial \xi _{i,l}}\equiv
J^{-1}\left( \frac{\partial J}{\partial \xi _{0,k}}\frac{\partial J}{%
\partial \xi _{i,l}}-\frac{\partial J}{\partial \xi _{0,l}}\frac{\partial J}{%
\partial \xi _{i,k}}\right)  \label{A.9}
\end{equation}
\begin{equation}
\frac{\partial J}{\partial \xi _{i,l}}\xi _{k,l}\equiv J\delta
_{k}^{i},\qquad \partial _{l}\frac{\partial ^{2}J}{\partial \xi
_{0,k}\partial \xi _{i,l}}\equiv 0  \label{A.10}
\end{equation}
one can test by direct substitution that the general solution of linear
equations (\ref{A.8}) has the form 
\begin{equation}
p_{k}=b_{0}\left( \partial _{k}\varphi +g^{\alpha }\left( \mathbf{\xi }%
\right) \partial _{k}\xi _{\alpha }\right) ,\qquad k=0,1,2,3  \label{A.11}
\end{equation}
where $b_{0}\neq 0$ is a constant, $g^{\alpha }\left( \mathbf{\xi }\right)
,\;\;\alpha =1,2,3$ are arbitrary functions of $\mathbf{\xi =}\left\{ \xi
_{1},\xi _{2},\xi _{3}\right\} $, and $\varphi $ is the dynamic variable $%
\xi _{0}$, which stops to be fictitious. Let us substitute (\ref{A.11}) in (%
\ref{A.7}). The term of the form $\partial J/\partial \xi _{0,k}\partial
_{k}\varphi $ is reduced to Jacobian and does not contribute to dynamic
equation. The terms of the form $\xi _{\alpha ,k}\partial J/\partial \xi
_{0,k}$ vanish due to identities (\ref{A.10}). We obtain 
\begin{equation}
\mathcal{A}\left[ \varphi ,\mathbf{\xi },\kappa ,j\right] =\int \left\{ -mcK%
\sqrt{g_{ik}j^{i}j^{k}}-j^{k}\pi _{k}\right\} d^{4}x\mathbf{,}  \label{A.12}
\end{equation}
where quantities $\pi _{k}$ are determined by the relations 
\begin{equation}
\pi _{k}=b_{0}\left( \partial _{k}\varphi +g^{\alpha }\left( \mathbf{\xi }%
\right) \partial _{k}\xi _{\alpha }\right) +\frac{e}{c}A_{k},\qquad k=0,1,2,3
\label{A.12b}
\end{equation}

Integration of (\ref{A.8}) in the form (\ref{A.11}) is that integration
which was mentioned in introduction as a conceptual operation which admits
to introduce a wave function. Note that coefficients in the system (\ref{A.8}%
) of equations for $p_{k}$ are constructed of minors of the Jacobian (\ref
{A.3}). It is the circumstance that admits to produce a general integration.

Variation of (\ref{A.12}) with respect to $\kappa ^{l}$ gives 
\begin{equation}
\frac{\delta \mathcal{A}}{\delta \kappa ^{l}}=-\frac{\lambda ^{2}mc\sqrt{%
g_{ik}j^{i}j^{k}}}{K}\kappa _{l}+\partial _{l}\frac{\lambda ^{2}mc\sqrt{%
g_{ik}j^{i}j^{k}}}{2K}=0  \label{A.12a}
\end{equation}
It can be written in the form 
\begin{equation}
\kappa _{l}=\partial _{l}\kappa =\frac{1}{2}\partial _{l}\ln \rho ,\qquad
e^{2\kappa }=\frac{\rho }{\rho _{0}}\equiv \frac{\sqrt{j_{s}j^{s}}}{\rho
_{0}mcK},  \label{A.13}
\end{equation}
where $\rho _{0}=$const is the integration constant. Substituting (\ref{A.2}%
) in (\ref{A.13}), we obtain dynamic equation for $\kappa $%
\begin{equation}
\hbar ^{2}\left( \partial _{l}\kappa \cdot \partial ^{l}\kappa +\partial
_{l}\partial ^{l}\kappa \right) =\frac{e^{-4\kappa }j_{s}j^{s}}{\rho _{0}^{2}%
}-m^{2}c^{2}  \label{A.14}
\end{equation}

Variation of (\ref{A.12}) with respect to $j^{k}$ gives 
\begin{equation}
\pi _{k}=-\frac{mcKj_{k}}{\sqrt{g_{ls}j^{l}j^{s}}}  \label{A.16}
\end{equation}
or 
\begin{equation}
\pi _{k}g^{kl}\pi _{l}=m^{2}c^{2}K^{2}  \label{A.15}
\end{equation}
Substituting the second equation (\ref{A.13}) in (\ref{A.16}), we obtain 
\begin{equation}
j_{k}=-\rho _{0}e^{2\kappa }\pi _{k},  \label{A.17}
\end{equation}

Now we eliminate the variables $j^{k}$ from the action (\ref{A.12}), using
relation (\ref{A.17}) and (\ref{A.13}). We obtain 
\begin{equation}
\mathcal{A}\left[ \varphi ,\mathbf{\xi },\kappa \right] =\int \rho
_{0}e^{2\kappa }\left\{ -m^{2}c^{2}K^{2}+\pi ^{k}\pi _{k}\right\} d^{4}x%
\mathbf{,}  \label{A.28}
\end{equation}
where $\pi _{k}$ is determined by the relation (\ref{A.12b}). Using
expression (\ref{A.2}) for $K,$ the first term of the action (\ref{A.28})
can be transformed as follows. 
\begin{eqnarray*}
-m^{2}c^{2}e^{2\kappa }K^{2} &=&-m^{2}c^{2}e^{2\kappa }\left( 1+\lambda
^{2}\left( \partial _{l}\kappa \partial ^{l}\kappa +\partial _{l}\partial
^{l}\kappa \right) \right) \\
&=&-m^{2}c^{2}e^{2\kappa }+\hbar ^{2}e^{2\kappa }\partial _{l}\kappa
\partial ^{l}\kappa -\frac{\hbar ^{2}}{2}\partial _{l}\partial
^{l}e^{2\kappa }
\end{eqnarray*}

Let us take into account that the last term has the form of divergence. It
does not contribute to dynamic equations and can be omitted. Omitting this
term, we obtain 
\begin{equation}
\mathcal{A}\left[ \varphi ,\mathbf{\xi },\kappa \right] =\int \rho
_{0}e^{2\kappa }\left\{ -m^{2}c^{2}+\hbar ^{2}\partial _{l}\kappa \partial
^{l}\kappa +\pi ^{k}\pi _{k}\right\} d^{4}x\mathbf{,}  \label{A.29}
\end{equation}

Instead of dynamic variables $\varphi ,\mathbf{\xi },\kappa $ we introduce $%
n $-component complex function 
\begin{equation}
\psi =\left\{ \psi _{\alpha }\right\} =\left\{ \sqrt{\rho }e^{i\varphi
}u_{\alpha }\left( \mathbf{\xi }\right) \right\} =\left\{ \sqrt{\rho _{0}}%
e^{\kappa +i\varphi }u_{\alpha }\left( \mathbf{\xi }\right) \right\} ,\qquad
\alpha =1,2,...n  \label{A.30}
\end{equation}
Here $u_{\alpha }$ are functions of only $\mathbf{\xi }=\left\{ \xi _{1},\xi
_{2},\xi _{3}\right\} $, having the following properties 
\begin{equation}
\sum\limits_{\alpha =1}^{\alpha =n}u_{\alpha }^{\ast }u_{\alpha }=1,\qquad -%
\frac{i}{2}\sum\limits_{\alpha =1}^{\alpha =n}\left( u_{\alpha }^{\ast }%
\frac{\partial u_{\alpha }}{\partial \xi _{\beta }}-\frac{\partial u_{\alpha
}^{\ast }}{\partial \xi _{\beta }}u_{\alpha }\right) =g^{\beta }\left( 
\mathbf{\xi }\right)  \label{A.31}
\end{equation}
where ($^{\ast }$) denotes complex conjugation. The number $n$ of components
of the wave function $\psi $ is chosen in such a way, that equations (\ref
{A.31}) have a solution. Then we obtain 
\begin{eqnarray}
\psi ^{\ast }\psi &\equiv &\sum\limits_{\alpha =1}^{\alpha =n}\psi _{\alpha
}^{\ast }\psi _{\alpha }=\rho =\rho _{0}e^{2\kappa },\qquad \partial
_{l}\kappa =\frac{\partial _{l}\left( \psi ^{\ast }\psi \right) }{2\psi
^{\ast }\psi }  \label{A.33} \\
\pi _{k} &=&-\frac{ib_{0}\left( \psi ^{\ast }\partial _{k}\psi -\partial
_{k}\psi ^{\ast }\cdot \psi \right) }{2\psi ^{\ast }\psi }+\frac{e}{c}%
A_{k},\qquad k=0,1,2,3  \label{A.34}
\end{eqnarray}
Substituting relations (\ref{A.33}), (\ref{A.34}) in (\ref{A.29}), we obtain
the action, written in terms of the wave function $\psi $%
\begin{eqnarray}
\mathcal{A}\left[ \psi ,\psi ^{\ast }\right] &=&\int \left\{ \left[ \frac{%
ib_{0}\left( \psi ^{\ast }\partial _{k}\psi -\partial _{k}\psi ^{\ast }\cdot
\psi \right) }{2\psi ^{\ast }\psi }-\frac{e}{c}A_{k}\right] \left[ \frac{%
ib_{0}\left( \psi ^{\ast }\partial ^{k}\psi -\partial ^{k}\psi ^{\ast }\cdot
\psi \right) }{2\psi ^{\ast }\psi }-\frac{e}{c}A^{k}\right] \right. 
\nonumber \\
&&+\left. \hbar ^{2}\frac{\partial _{l}\left( \psi ^{\ast }\psi \right)
\partial ^{l}\left( \psi ^{\ast }\psi \right) }{4\left( \psi ^{\ast }\psi
\right) ^{2}}-m^{2}c^{2}\right\} \psi ^{\ast }\psi d^{4}x  \label{A.35}
\end{eqnarray}

Let us use the identity 
\begin{eqnarray}
&&\frac{\left( \psi ^{\ast }\partial _{l}\psi -\partial _{l}\psi ^{\ast
}\cdot \psi \right) \left( \psi ^{\ast }\partial ^{l}\psi -\partial ^{l}\psi
^{\ast }\cdot \psi \right) }{4\psi ^{\ast }\psi }+\partial _{l}\psi ^{\ast
}\partial ^{l}\psi  \nonumber \\
&\equiv &\frac{\partial _{l}\left( \psi ^{\ast }\psi \right) \partial
^{l}\left( \psi ^{\ast }\psi \right) }{4\psi ^{\ast }\psi }+\frac{g^{ls}}{2}%
\psi ^{\ast }\psi \sum\limits_{\alpha ,\beta =1}^{\alpha ,\beta =n}Q_{\alpha
\beta ,l}^{\ast }Q_{\alpha \beta ,s}  \label{A.36}
\end{eqnarray}
where 
\begin{equation}
Q_{\alpha \beta ,l}=\frac{1}{\psi ^{\ast }\psi }\left| 
\begin{array}{cc}
\psi _{\alpha } & \psi _{\beta } \\ 
\partial _{l}\psi _{\alpha } & \partial _{l}\psi _{\beta }
\end{array}
\right| ,\qquad Q_{\alpha \beta ,l}^{\ast }=\frac{1}{\psi ^{\ast }\psi }%
\left| 
\begin{array}{cc}
\psi _{\alpha }^{\ast } & \psi _{\beta }^{\ast } \\ 
\partial _{l}\psi _{\alpha }^{\ast } & \partial _{l}\psi _{\beta }^{\ast }
\end{array}
\right|  \label{A.37}
\end{equation}
Then we obtain 
\begin{eqnarray}
\mathcal{A}\left[ \psi ,\psi ^{\ast }\right] &=&\int \left\{ \left(
ib_{0}\partial _{k}+\frac{e}{c}A_{k}\right) \psi ^{\ast }\left(
-ib_{0}\partial ^{k}+\frac{e}{c}A^{k}\right) \psi +\frac{b_{0}^{2}}{2}%
\sum\limits_{\alpha ,\beta =1}^{\alpha ,\beta =n}g^{ls}Q_{\alpha \beta
,l}Q_{\alpha \beta ,s}^{\ast }\psi ^{\ast }\psi \right.  \nonumber \\
&&\left. -m^{2}c^{2}\psi ^{\ast }\psi +\left( \hbar ^{2}-b_{0}^{2}\right) 
\frac{\partial _{l}\left( \psi ^{\ast }\psi \right) \partial ^{l}\left( \psi
^{\ast }\psi \right) }{4\psi ^{\ast }\psi }\right\} d^{4}x  \label{A.38}
\end{eqnarray}

Let us consider the case of irrotational flow, when $g^{\alpha }\left( 
\mathbf{\xi }\right) =0$ and the function $\psi $ has only one component. It
follows from (\ref{A.37}), that $Q_{\alpha \beta ,l}=0$, and only the last
term in (\ref{A.38}) is not bilinear with respect to $\psi ,\psi ^{\ast }$.
The constant $b_{0}$ is an arbitrary integration constant. One may set $%
b_{0}=\hbar $. Then we obtain instead of (\ref{A.38}) 
\begin{equation}
\mathcal{A}\left[ \psi ,\psi ^{\ast }\right] =\int \left\{ \left( i\hbar
\partial _{k}+\frac{e}{c}A_{k}\right) \psi ^{\ast }\left( -i\hbar \partial
^{k}+\frac{e}{c}A^{k}\right) \psi -m^{2}c^{2}\psi ^{\ast }\psi \right\}
d^{4}x  \label{A.39}
\end{equation}
Variation of the action (\ref{A.39}) with respect to $\psi ^{\ast }$
generates the Klein-Gordon equation 
\begin{equation}
\left( -i\hbar \partial _{k}+\frac{e}{c}A_{k}\right) \left( -i\hbar \partial
^{k}+\frac{e}{c}A^{k}\right) \psi -m^{2}c^{2}\psi =0  \label{A.40}
\end{equation}
Thus, description in terms of the Klein-Gordon equation is a special case of
the stochastic system description by means of the action (\ref{A.1}), (\ref
{A.2}).

In the case, when the fluid flow is rotational, and the wave function $\psi $
is two-component, the identity (\ref{A.36}) takes the form 
\begin{eqnarray}
&&\frac{\left( \psi ^{\ast }\partial _{l}\psi -\partial _{l}\psi ^{\ast
}\cdot \psi \right) \left( \psi ^{\ast }\partial ^{l}\psi -\partial ^{l}\psi
^{\ast }\cdot \psi \right) }{4\rho }-\frac{\left( \partial _{l}\rho \right)
\left( \partial ^{l}\rho \right) }{4\rho }  \nonumber \\
&\equiv &-\partial _{l}\psi ^{\ast }\partial ^{l}\psi +\frac{1}{4}\left(
\partial _{l}s_{\alpha }\right) \left( \partial ^{l}s_{\alpha }\right) \rho
\label{A.41}
\end{eqnarray}
where 3-vector $\mathbf{s=}\left\{ s_{1},s_{2},s_{3},\right\} $ is defined
by the relation 
\begin{equation}
\rho =\psi ^{\ast }\psi ,\qquad s_{\alpha }=\frac{\psi ^{\ast }\sigma
_{\alpha }\psi }{\rho },\qquad \alpha =1,2,3  \label{A.42}
\end{equation}
\begin{equation}
\psi =\left( _{\psi _{2}}^{\psi _{1}}\right) ,\qquad \psi ^{\ast }=\left(
\psi _{1}^{\ast },\psi _{2}^{\ast }\right) ,  \label{A.43}
\end{equation}
and Pauli matrices $\mathbf{\sigma }=\left\{ \sigma _{1},\sigma _{2},\sigma
_{3}\right\} $ have the form 
\begin{equation}
\sigma _{1}=\left( 
\begin{array}{cc}
0 & 1 \\ 
1 & 0
\end{array}
\right) ,\qquad \sigma _{2}=\left( 
\begin{array}{cc}
0 & -i \\ 
i & 0
\end{array}
\right) ,\qquad \sigma _{1}=\left( 
\begin{array}{cc}
1 & 0 \\ 
0 & -1
\end{array}
\right)  \label{A.44}
\end{equation}
Note that 3-vectors $\mathbf{s}$ and $\mathbf{\sigma }$ are vectors in the
space $V_{\xi }$ of the Clebsch potentials $\mathbf{\xi }=\left\{ \xi
_{1},\xi _{2},\xi _{3}\right\} $ and transform as vectors at the
transformations (\ref{a0.6}) of Clebsch potentials $\mathbf{\xi }$. The wave
function $\psi $ is a spinor in the space $V_{\xi }$. The quantities $%
\mathbf{s}$, $\mathbf{\sigma }$ and $\psi $ are scalars in the 3-space $%
V_{x} $ of usual coordinates $\mathbf{x}=\left\{ x^{1},x^{2},x^{3}\right\} $.

In general, transformations of Clebsch potentials $\mathbf{\xi }$ and those
of coordinates $\mathbf{x}$ are independent. However, the action (\ref{A.35}%
) does not contain any reference to the Clebsch potentials $\mathbf{\xi }$
and transformations (\ref{a0.6}) of $\mathbf{\xi }$. If we consider only
linear transformations of space coordinates $\mathbf{x}$%
\begin{equation}
x^{\alpha }\rightarrow \tilde{x}^{\alpha }=b^{\alpha }+\omega _{.\beta
}^{\alpha }x^{\beta },\qquad \alpha =1,2,3  \label{A.44a}
\end{equation}
nothing prevents from accompanying any transformation (\ref{A.44a}) with the
similar transformation 
\begin{equation}
\xi _{\alpha }\rightarrow \tilde{\xi}_{\alpha }=b^{\alpha }+\omega _{.\beta
}^{\alpha }\xi _{\beta },\qquad \alpha =1,2,3  \label{A.44b}
\end{equation}
of Clebsch potentials $\mathbf{\xi }$. The formulas for linear
transformation of vectors and spinors in $V_{x}$ do not contain the
coordinates $\mathbf{x}$ explicitly, and one can consider vectors and
spinors in $V_{\xi }$ as vectors and spinors in $V_{x}$, provided we
consider linear transformations (\ref{A.44a}), (\ref{A.44b}) always together.

Using identity (\ref{A.41}), we obtain from (\ref{A.35}) 
\begin{equation}
\mathcal{A}\left[ \psi ,\psi ^{\ast }\right] =\int \left\{ \left( i\hbar
\partial _{k}+\frac{e}{c}A_{k}\right) \psi ^{\ast }\left( -i\hbar \partial
^{k}+\frac{e}{c}A^{k}\right) \psi -m^{2}c^{2}\rho -\frac{\hbar ^{2}}{4}%
\left( \partial _{l}s_{\alpha }\right) \left( \partial ^{l}s_{\alpha
}\right) \rho \right\} d^{4}x  \label{A.45}
\end{equation}
Dynamic equation, generated by the action (\ref{A.45}), has the form 
\begin{eqnarray}
&&\left( -i\hbar \partial _{k}+\frac{e}{c}A_{k}\right) \left( -i\hbar
\partial ^{k}+\frac{e}{c}A^{k}\right) \psi -\left( m^{2}c^{2}+\frac{\hbar
^{2}}{4}\left( \partial _{l}s_{\alpha }\right) \left( \partial ^{l}s_{\alpha
}\right) \right) \psi  \nonumber \\
&=&\hbar ^{2}\frac{\partial _{l}\left( \rho \partial ^{l}s_{\alpha }\right) 
}{2\rho }\left( \sigma _{\alpha }-s_{\alpha }\right) \psi  \label{A.46}
\end{eqnarray}

The gradient of the unit 3-vector $\mathbf{s}=\left\{
s_{1},s_{2},s_{3}\right\} $ describes rotational component of the fluid
flow. If $\mathbf{s}=$const the dynamic equation (\ref{A.46}) turns to the
conventional Klein-Gordon equation (\ref{A.40}). Curl of the vector field $%
\pi _{k}$, determined by the relation (\ref{A.34}), is expressed only via
derivatives of the unit 3-vector $\mathbf{s}$.

To show this, let us represent the wave function (\ref{A.30}) in the form 
\begin{equation}
\psi =\sqrt{\rho }e^{i\varphi }\left( \mathbf{n\sigma }\right) \chi ,\qquad
\psi ^{\ast }=\sqrt{\rho }e^{-i\varphi }\chi ^{\ast }\left( \mathbf{\sigma n}%
\right) ,\qquad \mathbf{n}^{2}=1,\qquad \chi ^{\ast }\chi =1  \label{A.47}
\end{equation}
where $\mathbf{n}=\left\{ n_{1},n_{2},n_{3}\right\} $ is some unit 3-vector, 
$\chi =\left( _{\chi _{2}}^{\chi _{1}}\right) $, $\chi ^{\ast }=\left( \chi
_{1}^{\ast },\chi _{2}^{\ast }\right) $ are constant two-component
quantities, and $\mathbf{\sigma }=\left\{ \sigma _{1},\sigma _{2},\sigma
_{3}\right\} $ are Pauli matrices (\ref{A.44}). The unit vector $\mathbf{s}$
and the unit vector $\mathbf{n}$ are connected by means of the relations 
\begin{equation}
\mathbf{s}=2\mathbf{n}\left( \mathbf{nz}\right) -\mathbf{z},\qquad \mathbf{n}%
=\frac{\mathbf{s}+\mathbf{z}}{\sqrt{2\left( 1+\left( \mathbf{sz}\right)
\right) }}  \label{A.48}
\end{equation}
where $\mathbf{z}$ is a constant unit vector defined by the relation 
\begin{equation}
\mathbf{z}=\chi ^{\ast }\mathbf{\sigma }\chi ,\qquad \mathbf{z}^{2}=\chi
^{\ast }\chi =1  \label{A.49}
\end{equation}
All 3-vectors $\mathbf{n}$, $\mathbf{s}$, $\mathbf{z}$ are vectors in $%
V_{\xi }$. Let us substitute the relation (\ref{A.47}) into expression $%
\partial _{l}\pi _{k}-\partial _{k}\pi _{l}$ for the curl of the vector
field $\pi _{k}$ defined by the relation (\ref{A.34}). Then gradually
reducing powers of $\sigma $ by means of the identity 
\begin{equation}
\sigma _{\alpha }\sigma _{\beta }\equiv \delta _{\alpha \beta }+i\varepsilon
_{\alpha \beta \gamma }\sigma _{\gamma },\qquad \alpha ,\beta =1,2,3
\label{A.50}
\end{equation}
where $\varepsilon _{\alpha \beta \gamma }$ is the Levi-Chivita pseudotensor
($\varepsilon _{123}=1$), we obtain after calculations 
\begin{eqnarray}
\pi _{k} &=&-\frac{ib_{0}\left( \psi ^{\ast }\partial _{k}\psi -\partial
_{k}\psi ^{\ast }\cdot \psi \right) }{2\psi ^{\ast }\psi }+\frac{e}{c}A_{k} 
\nonumber \\
&=&b_{0}\left( \partial _{k}\varphi +\varepsilon _{\alpha \beta \gamma
}n_{\alpha }\partial _{k}n_{\beta }z_{\gamma }\right) +\frac{e}{c}A_{k}
\qquad k=0,1,2,3  \label{A.51}
\end{eqnarray}
\begin{equation}
\partial _{k}\pi _{l}-\partial _{l}\pi _{k}=-4b_{0}\left[ \partial _{k}%
\mathbf{n}\times \partial _{l}\mathbf{n}\right] \mathbf{z}+\frac{e}{c}\left(
\partial _{k}A_{l}-\partial _{l}A_{k}\right) ,\qquad k,l=0,1,2,3
\label{A.52}
\end{equation}
The relation (\ref{A.52}) may be expressed also via the 3-vector $\mathbf{s}$%
, provided we use the formulae (\ref{A.48}).

Note that the two-component form of the wave function can describe
irrotational flow. For instance, if $\psi =\left( _{\psi _{1}}^{\psi
_{1}}\right) $, $s_{1}=1,$ $s_{2}=s_{3}=0$, the dynamic equation (\ref{A.46}%
) reduces to the form (\ref{A.40}), and curl of $\pi _{k}$, defined by (\ref
{A.52}) vanishes.

\section{Interpretation of solutions of the free Klein-Gordon equation}

Let us write the action (\ref{A.12}), (\ref{A.2}) in the arbitrary
curvilinear coordinate system. We obtain for dynamic system $\mathcal{S}_{%
\mathrm{KG}}$ 
\begin{eqnarray}
\mathcal{A}\left[ \varphi ,\mathbf{\xi },\kappa ,j\right] &=&\int \left\{
-mcK\sqrt{g_{ik}j^{i}j^{k}}-j^{k}\pi _{k}\right\} \sqrt{-g}d^{4}x\mathbf{,}
\label{a4.1} \\
K &=&\sqrt{1+\lambda ^{2}\left( g_{ik}\kappa ^{i}\kappa ^{k}+\frac{1}{\sqrt{%
-g}}\partial _{l}\left( \sqrt{-g}\kappa ^{l}\right) \right) },\qquad \lambda
=\frac{\hbar }{mc}  \label{a4.2}
\end{eqnarray}
\begin{equation}
g=\det \left| \left| g_{ik}\right| \right|  \label{a4.2a}
\end{equation}
where quantities $\pi _{k}$ are determined by the relations (\ref{A.12b}) 
\begin{equation}
\pi _{k}=b_{0}\left( \partial _{k}\varphi +g^{\alpha }\left( \mathbf{\xi }%
\right) \partial _{k}\xi _{\alpha }\right) +\frac{e}{c}A_{k},\qquad k=0,1,2,3
\label{a4.3}
\end{equation}
Trying to interpret dynamic system $\mathcal{S}_{\mathrm{KG}}$ described by
the action (\ref{a4.1}) -- (\ref{a4.3}), we shall \textit{use only dynamic
considerations}. This means that $\mathcal{S}_{\mathrm{KG}}$ is considered
to be a fluid, and world lines of the fluid particles are interpreted as
average world lines of stochastic particles. We consider $\mathcal{S}_{%
\mathrm{KG}}$ as a fluid and interpret its motion in terms of hydrodynamics.
We ignore interpretation, based on quantum principles, because this
interpretation in terms of the wave function is purely empirical. It has
been tested only for nonrelativistic quantum phenomena. In the relativistic
case there are problems for interpretation of expressions constructed on the
basis of the wave function.

From viewpoint of hydrodynamic interpretation the wave function is only a
very special method of description of $\mathcal{S}_{\mathrm{KG}}$. This
method is very convenient for solution of dynamic equations, because they
are linear in terms of wave function, but it is not the best method for
interpretation of the fluid motion.

We shall interpret dynamic system, following the behaviour of such physical
quantities as the current $j^{i}$ and the energy-momentum tensor $T^{ik}$.
The energy-momentum tensor for $\mathcal{S}_{\mathrm{KG}}$ is obtained by
means of relation (\ref{a2.2}). Applying the relation (\ref{a2.2}) to the
action (\ref{a4.1}) -- (\ref{a4.3}) and taking into account that the
Lagrangian density vanishes due to (\ref{A.16}), we obtain 
\begin{eqnarray*}
T^{ik} &=&\frac{mcKj^{i}j^{k}}{\sqrt{g_{ls}j^{l}j^{s}}}+\lambda ^{2}\frac{mc%
\sqrt{g_{ls}j^{l}j^{s}}}{K}\kappa ^{i}\kappa ^{k}+\lambda ^{2}\frac{mc\sqrt{%
g_{ls}j^{l}j^{s}}}{K}\left( -\frac{1}{2\left( \sqrt{-g}\right) }%
g^{ik}\partial _{l}\left( \sqrt{-g}\kappa ^{l}\right) \right)  \\
&&-\lambda ^{2}\left( \frac{1}{2}g^{ik}\kappa ^{j}\right) \partial _{j}\frac{%
mc\sqrt{g_{ls}j^{l}j^{s}}}{K}
\end{eqnarray*}
We use the  relation (\ref{A.13}) 
\begin{equation}
e^{2\kappa }=\frac{\rho }{\rho _{0}}\equiv \frac{\sqrt{j_{s}j^{s}}}{\rho
_{0}mcK}.  \label{a4.8}
\end{equation}
Returning to the inertial coordinate system, where $\sqrt{-g}=c$, we obtain 
\begin{equation}
T^{ik}=\frac{j^{i}j^{k}}{\rho }+\hbar ^{2}\rho \kappa ^{i}\kappa ^{k}-\frac{%
\hbar ^{2}}{2}g^{ik}\partial _{l}\left( \rho \kappa ^{l}\right) 
\label{a4.7}
\end{equation}
or
\begin{eqnarray}
T^{ik} &=&\frac{j^{i}j^{k}+j_{\mathrm{st}}^{i}j_{\mathrm{st}}^{k}}{\rho }-%
\frac{\hbar }{2}g^{ik}\partial _{l}j_{\mathrm{st}}^{l},  \label{a4.10} \\
j_{\mathrm{st}}^{k} &=&\hbar \rho \kappa ^{k}=\frac{\hbar }{2}\partial
^{k}\rho ,\qquad \rho =\rho _{0}e^{2\kappa }=\frac{\sqrt{j^{l}j_{l}}}{mcK}%
,\qquad \kappa ^{i}=g^{ik}\partial _{k}\kappa   \label{a4.10a}
\end{eqnarray}
The energy-momentum tensor (\ref{a4.10}) looks as the energy-momentum tensor
of two fluids. Flux of one fluid is described by the 4-vector $j^{i}$,
whereas the flux of other fluid is described by the 4-vector $j_{\mathrm{st}%
}^{i}=\hbar \rho \kappa ^{i}=\hbar g^{ik}\rho \partial _{k}\kappa $. In the
first fluid there is no pressure. The pressure $p$ in the second fluid is
described by the expression 
\[
p=\frac{\hbar }{2}\partial _{l}j_{\mathrm{st}}^{l}
\]
The two fluids interact via common factor $\rho =\rho _{0}e^{2\kappa }$.

Eigenvectors $u_{i}$ and eigenvalues $\mu $ of the energy-momentum tensor
are defined by the relations 
\begin{equation}
T^{ik}u_{k}=\mu g^{ik}u_{k},\qquad u_{k}=\alpha j_{k}+\beta j_{\mathrm{st}k}
\label{a4.11}
\end{equation}
Calculation gives 
\begin{equation}
\mu _{1,2}=-\frac{\hbar }{2}\partial _{l}j_{\mathrm{st}}^{l}+\frac{1}{2\rho }%
\left( j_{l}j^{l}+j_{\mathrm{st}l}j_{\mathrm{st}}^{l}\pm \sqrt{\left(
j^{l}j_{l}-\left( j_{\mathrm{st}}^{l}j_{\mathrm{st}l}\right) \right)
^{2}+4\left( j_{\mathrm{st}}^{l}j_{l}\right) ^{2}}\right)  \label{a4.12}
\end{equation}
\begin{equation}
\beta _{1,2}=\frac{\left( j_{l}j^{l}-j_{\mathrm{st}l}j_{\mathrm{st}}^{l}\pm 
\sqrt{\left( j^{l}j_{l}-\left( j_{\mathrm{st}}^{l}j_{\mathrm{st}l}\right)
\right) ^{2}+4\left( j_{\mathrm{st}}^{l}j_{l}\right) ^{2}}\right) }{2j_{l}j_{%
\mathrm{st}}^{l}}\alpha  \label{a4.12a}
\end{equation}

There is a very simple and interesting case, when one of fluids vanishes and 
\begin{equation}
j^{i}\equiv 0,\qquad i=0,1,2,3.  \label{a4.14}
\end{equation}
As it follows from (\ref{A.14}), in this case the scalar $e^{\kappa }$
satisfies the free Klein-Gordon equation 
\begin{equation}
\left( \hbar ^{2}\partial _{l}\partial ^{l}+m^{2}c^{2}\right) e^{\kappa }=0
\label{a4.15}
\end{equation}
and 
\begin{equation}
K=\sqrt{1+\lambda ^{2}e^{-\kappa }\partial _{l}\partial ^{l}e^{\kappa }}=0
\label{a4.16}
\end{equation}
On the other side, in this case the wave function (\ref{A.30}) is
one-component, and it satisfies the Klein-Gordon equation (\ref{A.40}). If
the condition (\ref{a4.14}) is satisfied, the wave function has the form $%
\psi =e^{\kappa +i\varphi }$, $\varphi =$const. Equations (\ref{a4.15}) and (%
\ref{A.40}) are compatible, provided $A_{i}=0$.

Under conditions (\ref{a4.14}) and (\ref{a4.16}) the density $\rho $ (\ref
{a4.8}) and the energy-momentum tensor (\ref{a4.10}) become indefinite.
Besides, $\kappa $ is to be real and $e^{\kappa }\geq 0$.

To obtain the canonical energy-momentum tensor for the irrotational flow, we
consider the action of the form (\ref{A.39}) with additional condition $%
A_{i}=0$, where $\psi $ is the one-component wave function. The action has
the form 
\begin{equation}
\mathcal{A}\left[ \psi ,\psi ^{\ast }\right] =\int \left\{ \hbar
^{2}\partial _{k}\psi ^{\ast }\cdot \partial ^{k}\psi -m^{2}c^{2}\psi ^{\ast
}\psi \right\} d^{4}x  \label{a4.17}
\end{equation}
Then canonical energy-momentum tensor 
\begin{equation}
T_{\mathrm{c}}\,_{k}^{l}=\hbar ^{2}\left( \partial _{k}\psi ^{\ast }\cdot
\partial ^{l}\psi +\partial ^{l}\psi ^{\ast }\cdot \partial _{k}\psi \right)
-\delta _{k}^{l}\left( \hbar ^{2}\partial _{k}\psi ^{\ast }\cdot \partial
^{k}\psi -m^{2}c^{2}\psi ^{\ast }\psi \right)   \label{a4.18}
\end{equation}
coincides with the relation (\ref{a4.10}).

Indeed, let us substitute in (\ref{a4.18}) the expression 
\begin{equation}
\partial _{k}\psi =\sqrt{\rho _{0}}e^{\kappa }\partial _{k}\kappa -\frac{i}{%
\hbar \sqrt{\rho _{0}}}e^{-\kappa }j_{k}=\sqrt{\rho }\partial _{k}\kappa -%
\frac{i}{\hbar \sqrt{\rho }}j_{k}  \label{a4.19a}
\end{equation}
which is valid for the irrotational flow. 
\[
T_{\mathrm{c}kl}=\frac{j_{l}j_{k}}{\rho }+\hbar ^{2}\rho \partial _{l}\kappa
\cdot \partial _{k}\kappa -\rho _{0}e^{2\kappa }g_{kl}\left( \hbar
^{2}\partial _{s}\kappa \cdot \partial ^{s}\kappa +\frac{1}{\rho _{0}^{2}}%
e^{-4\kappa }j_{s}j^{s}-m^{2}c^{2}\right) 
\]
Taking into account dynamic equation (\ref{A.14}), we eliminate two last
terms 
\begin{equation}
T_{\mathrm{c}kl}=\frac{j_{l}j_{k}}{\rho }+\hbar ^{2}\rho \partial _{l}\kappa
\cdot \partial _{k}\kappa -g_{kl}\frac{\hbar ^{2}}{2}\partial _{s}\partial
^{s}e^{2\kappa }  \label{a4.19b}
\end{equation}
and the relations (\ref{a4.7}) and (\ref{a4.19b}) coincide.

Solution of equation (\ref{a4.15}), which is stationary in the whole
space-time, is $e^{\kappa }=e^{\kappa \left( \mathbf{x}\right) }$. It
satisfies the equation 
\[
\left( \hbar ^{2}\partial _{\alpha }\partial _{\alpha }-m^{2}c^{2}\right)
e^{\kappa \left( \mathbf{x}\right) }=0 
\]
Solutions of this equation have the form 
\begin{equation}
\psi _{\mathbf{k}}=e^{\kappa \left( \mathbf{x}\right) }=B_{\mathbf{k}}\cos
\left( \frac{\mathbf{kx}}{\hbar }+\phi _{\mathbf{k}}\right) ,\qquad \mathbf{k%
}^{2}=m^{2}c^{2},\qquad B_{\mathbf{k}}=\text{const}  \label{a4.20}
\end{equation}
None of these solutions satisfies the condition $\psi \geq 0$ except for $%
\psi =0$. For this solution 
\[
T^{00}=0,\qquad T^{0\alpha }=0, 
\]
and the solution $\psi =0$ may be interpreted as a vacuum state. Possibility
of construction of a linear combination of solutions (\ref{a4.20}) which
would be positive in the whole space-time seems to be problematic, although
it is possible in small regions $\Omega _{\mathrm{v}}$ of the space-time. At
the boundaries of $\Omega _{\mathrm{v}}$ the solution $\psi $ becomes
negative and $\kappa $ becomes to be complex. It corresponds to $j^{k}\neq 0$%
, i.e. to appearance of SWLs (particles and antiparticles). For solutions $%
e^{\kappa }$, which are linear combinations of solutions (\ref{a4.20}), $%
T^{00}\geq 0$, and these solutions may be interpreted as excited states of
vacuum inside the region $\Omega _{\mathrm{v}}$. At the boundaries of $%
\Omega _{\mathrm{v}}$ pairs of SWLs appear. In general, the pairs are
virtual in the sense that for each of SWL $p_{i}p^{i}\neq m^{2}c^{2}$.

Another case, when there is only one fluid, described by the vector $j^{k}$
and $j_{\mathrm{st}}^{l}=0$ corresponds to description of statistical
ensemble of relativistic classical (deterministic) particles.

\section{Interpretation of $\protect\kappa $-field in the Klein-Gordon
dynamic system}

Let us consider solutions of the free Klein-Gordon equation ($A_{k}=0$, \ $%
k=0,1,2,3$). For simplicity of calculations we set inside this paragraph $%
\hbar =c=1$. The general solution of the free Klein-Gordon equation has the
form 
\begin{eqnarray}
\psi \left( t,\mathbf{x}\right) &=&\left( 2\pi \right) ^{-3/2}\int
e^{-i\varepsilon _{k}E\left( \mathbf{k}\right) t+i\mathbf{kx}}\frac{a\left(
K\right) }{\sqrt{2E\left( \mathbf{k}\right) }}dK,\qquad  \label{a7.1} \\
K &=&\left\{ \varepsilon _{k},\mathbf{k}\right\} ,\qquad E\left( \mathbf{k}%
\right) =\left| \sqrt{m^{2}+\mathbf{k}^{2}}\right| ,\qquad \mathbf{kx}\equiv
-k_{\alpha }x^{\alpha }\equiv k_{\alpha }x_{\alpha }  \label{a7.2}
\end{eqnarray}
where $a\left( K\right) $ is a function of $K=\left\{ \varepsilon _{k},%
\mathbf{k}\right\} $, $\;\mathbf{k}=\left\{ k_{1},k_{2},k_{3}\right\} $,$%
\;\;\varepsilon _{k}=\pm 1$ and 
\begin{equation}
\int \left( .\right) dK\equiv \sum\limits_{\varepsilon _{k}=\pm 1}\int
\left( .\right) d\mathbf{k}\equiv \sum\limits_{\varepsilon _{k}=\pm
1}\iiint_{-\infty }^{+\infty }\left( .\right) dk_{1}dk_{2}dk_{3}
\label{a7.3}
\end{equation}
Let us set 
\begin{equation}
a_{P}\left( K\right) =a_{\varepsilon _{p},\mathbf{p}}\left( K\right) =A\sqrt{%
2E\left( \mathbf{k}\right) }\exp \left( -\frac{\left( \mathbf{k}-\mathbf{p}%
\right) ^{2}}{2\Delta }\right) \delta _{\varepsilon _{k},\varepsilon
_{p}},\qquad A=\text{const}  \label{a7.4}
\end{equation}
The quantities $P=\left\{ \varepsilon _{p},\mathbf{p}\right\} $ and $\Delta $
are parameters of the solution. We suppose 
\begin{equation}
\sqrt{\Delta }\ll \lambda ^{-1}=m,\qquad \hbar =c=1  \label{a7.6}
\end{equation}
Under the condition (\ref{a7.6}) the following approximate expression for
the wave function is obtained 
\[
\psi _{P}\left( t,\mathbf{x}\right) =\frac{B\Delta ^{3/2}}{\left(
1+i\varepsilon _{k}\Delta t\left( \frac{3}{2E}+\frac{\mathbf{v}^{2}}{2}%
\right) \right) }\exp \left( -\frac{\Delta \left( \mathbf{x}-\mathbf{v}%
t\right) ^{2}}{2}-i\varepsilon _{p}E\left( \mathbf{p}\right) t+i\mathbf{px}%
\right) 
\]
where 
\begin{equation}
\mathbf{v}=\frac{\varepsilon _{p}\mathbf{p}}{E\left( \mathbf{p}\right) }%
,\qquad B=B^{\ast }=\text{const}  \label{a7.7}
\end{equation}
At $\left| t\right| \ll E/\Delta $ we shall use a simpler expression for the
wave function 
\begin{equation}
\psi _{P}\left( t,\mathbf{x}\right) =B\Delta ^{3/2}\exp \left( -\frac{\Delta
\left( \mathbf{x}-\mathbf{v}t\right) ^{2}}{2}-i\varepsilon _{p}E\left( 
\mathbf{p}\right) t+i\mathbf{px}\right)  \label{a7.8}
\end{equation}
In the space-time region $\left| \mathbf{x}-\mathbf{v}t\right| <\Delta
^{-1/2}$ we obtain the following values for the quantities $j^{k}$, $\kappa
^{k}$, $T_{k}^{0}$ 
\begin{equation}
\begin{array}{lll}
j^{0}=\left( \varepsilon _{p}E\left( \mathbf{p}\right) +O\left( \Delta
\right) \right) \rho & \kappa ^{0}=\Delta \left( \mathbf{x}-\mathbf{v}%
t\right) \mathbf{v} & T_{0}^{0}=\left( E^{2}\left( \mathbf{p}\right)
+O\left( \Delta \right) \right) \rho \\ 
j^{\alpha }=\left( p_{\alpha }+O\left( \Delta \right) \right) \rho & \kappa
^{\alpha }=\Delta \left( x^{\alpha }-v^{\alpha }t\right) & T_{\alpha
}^{0}=\left( \varepsilon _{p}E\left( \mathbf{p}\right) p_{a}+O\left( \Delta
\right) \right) \rho \\ 
\rho =\psi _{P}^{\ast }\psi _{P} & m_{\mathrm{eff}}^{2}=m^{2}+O\left( \Delta
\right) & T_{\alpha }^{\beta }=\left( p_{\alpha }p^{\beta }+O\left( \Delta
\right) \right) \rho
\end{array}
\label{a7.9}
\end{equation}
where 
\begin{equation}
\psi _{P}^{\ast }\psi _{P}=\rho _{0}e^{2\kappa }=B^{2}\Delta ^{3}\exp \left(
-\Delta \left( \mathbf{x}-\mathbf{v}t\right) ^{2}\right)  \label{a7.11}
\end{equation}

Interpretation from hydrodynamic viewpoint looks as follows. We assume that
the state $P=\left\{ 1,\mathbf{p}\right\} $ with $\varepsilon _{p}=1$
corresponds to a particle, and the state $P=\left\{ -1,\mathbf{p}\right\} $
with $\varepsilon _{p}=-1$ corresponds to an antiparticle, because as it
follows from (\ref{a7.9}) the flux density $j^{0}$ has different sign in the
two states. The total number $N$ of SWLs is 
\begin{equation}
N_{P}=\varepsilon _{p}E\left( \mathbf{p}\right) I,\qquad I=\int \psi
_{P}^{\ast }\psi _{P}d\mathbf{x}=B^{2}\left( \pi \Delta \right) ^{3/2}
\label{a7.12}
\end{equation}
The total energy-momentum vector 
\begin{equation}
P_{k}=\left\{ E^{2}\left( \mathbf{p}\right) I,E\left( \mathbf{p}\right) 
\mathbf{p}I\right\}  \label{a7.13}
\end{equation}
does not depend on orientation $\varepsilon _{p}$, whereas the energy per
one SWL 
\begin{equation}
\frac{P_{0}}{N_{P}}=\frac{E_{P}}{N_{P}}=\varepsilon _{p}E\left( \mathbf{p}%
\right)  \label{a7.14}
\end{equation}
depends on $\varepsilon _{p}$. It is negative for $\varepsilon _{p}=-1$,
because the number $N_{P}$ of SWLs is negative.\ Thus, the state $\psi _{P}$
describes a fluid, consisting of a cloud of $N_{P}$ SWLs, moving with
velocity $\mathbf{v}=\mathbf{p}/E\left( \mathbf{p}\right) +O\left( \Delta
\right) $. The pressure in the fluid is small (of the order $O\left( \Delta
\right) $), and one can speak on freely moving SWLs (particles or
antiparticles).

From viewpoint of quantum principles the state $\psi _{\left\{ 1,\mathbf{p}%
\right\} }$ (for particles) is interpreted similarly. But for interpretation
of the state $\psi _{\left\{ -1,\mathbf{p}\right\} }$ (for antiparticles)
there are problems \cite{FV58}, connected with negative value of the
quantity $E_{P}/N_{P}$ (energy per SWL).

Now let us consider the state $\psi =\psi _{\left\{ 1,\mathbf{p}\right\}
}+\psi _{\left\{ 1,-\mathbf{p}\right\} }$. In the conventional quantum
mechanics the wave function is a fundamental object of theory. The wave
functions $\psi _{\left\{ 1,\mathbf{p}\right\} }$ and $\psi _{\left\{ 1,-%
\mathbf{p}\right\} }$ evolve independently and this fact is interpreted in
the sense that the particles described by $\psi _{\left\{ 1,\mathbf{p}%
\right\} }$ and the particles described by $\psi _{\left\{ 1,-\mathbf{p}%
\right\} }$ do not interact. Thus, from viewpoint of quantum principles the
state $\psi $ describes two clouds of particles, moving with velocities $%
\mathbf{v}$ and $-\mathbf{v}$ one through another without interaction.

From hydrodynamic viewpoint the state $\psi =\psi _{\left\{ 1,\mathbf{p}%
\right\} }+\psi _{\left\{ 1,-\mathbf{p}\right\} }$ describes rather
complicated picture of two colliding clouds of particles. According to the
hydrodynamic interpretation the world lines of particles in the
two-dimensional space-time are defined by the equation 
\begin{equation}
\frac{dx}{dt}=\frac{j^{1}\left( t,x\right) }{j^{0}\left( t,x\right) }=\frac{%
\left( p\sinh \left( 2\Delta xvt\right) \right) -\Delta vt\left( \sin \left(
2px\right) \right) }{E\left( \cosh \left( 2\Delta xvt\right) +\cos \left(
2px\right) \right) +\Delta xv\left( \sin \left( 2px\right) \right) }+O\left(
\Delta ^{2}\right)  \label{a7.17}
\end{equation}
Shape of world lines, obtained by numerical calculations for $-2/\sqrt{%
\Delta }<t<2/\sqrt{\Delta }$ is shown in figure 1. One can see, that after
collision of two clouds the particles of clouds reflect and move in the
opposite direction. This situation correlates with values of the
energy-momentum tensor and other hydrodynamic quantities, calculated in the
space-time region $x^{2},t^{2}\le \Delta ^{-1}$, where the two clouds
overlap. These quantities are calculated to within $\Delta ^{0}$ 
\begin{equation}
\begin{array}{lll}
j^{0}=2CE\left( p\right) \cos ^{2}\left( px\right) & \kappa _{0}=0 & 
T^{00}=2CE^{2}\left( p\right) \cos ^{2}\left( px\right) \\ 
j_{1}=0 & \kappa _{1}=-p\tan \left( px\right) & T^{01}=0 \\ 
\rho =2C\cos ^{2}\left( px\right) & m_{\mathrm{eff}}^{2}=E^{2}\left( p\right)
& T^{11}=2Cp^{2}\sin ^{2}\left( px\right)
\end{array}
\label{a7.15}
\end{equation}
where 
\begin{equation}
C=B^{2}\exp \left( -\Delta \left( x^{2}+v^{2}t^{2}\right) \right)
\label{a7.16}
\end{equation}
The tensor energy-momentum component $T^{00}=j^{0}j^{0}/\rho $ is determined
by the 4-vector $j^{k}$, whereas the pressure in the fluid is determined by
the component $T^{11}=\rho \kappa ^{1}\kappa ^{1}$.

If the particles of two clouds do not interact, the energy-momentum tensor
is a sum of energy-momentum tensors for each cloud separately. It has the
form 
\begin{equation}
T^{00}=2CE^{2}\left( p\right) ,\qquad T^{01}=0,\qquad T^{11}=2Cp^{2}
\label{a7.16a}
\end{equation}
and this expression distinguishes from (\ref{a7.15})

Let us consider the state $\psi =\psi _{\left\{ \varepsilon _{p},\mathbf{p}%
\right\} }+\psi _{\left\{ -\varepsilon _{p},\mathbf{p}\right\} }$. From
viewpoint of quantum principles the state $\psi $ describes a cloud of
particles and a cloud of antiparticles, moving with velocities $\mathbf{v}$
and $-\mathbf{v}$ one through another without interaction.

From viewpoint of hydrodynamic interpretation the state $\psi _{\left\{ 1,%
\mathbf{p}\right\} }+\psi _{\left\{ -1,\mathbf{p}\right\} }$ describes a
complicated picture of interacting particles and antiparticles. In the
process of interaction the pair production and the pair annihilation take
place. The process of interaction is balanced, and finally we see a cloud of
particles and a cloud of antiparticles escaping one from another, as if they
have passed one through another without interaction.

In the two-dimensional space-time the mean world lines are described by the
equation 
\[
\frac{dx}{dt}=\frac{j^{1}}{j^{0}}=\frac{p\cosh \left( 2\Delta xvt\right)
+p\cos \left( 2Et\right) -\Delta vt\sin \left( 2Et\right) }{E\sinh \left(
2\Delta xvt\right) +\Delta xv\sin \left( 2Et\right) } 
\]
Shape of world lines obtained by numerical calculations for $-2/\sqrt{\Delta 
}<x<2/\sqrt{\Delta }$ is shown in figures 2. In the state $\psi _{\left\{ 1,%
\mathbf{p}\right\} }+\psi _{\left\{ -1,\mathbf{p}\right\} }$ colliding
particles and antiparticles annihilate, their energy transforms to energy of
the $\kappa $-field. Thereafter the $\kappa $-field generate SWL pairs and
energy of the $\kappa $-field turns to energy of escaping particles and
antiparticles. Distribution of space-time regions, where effective mass $m_{%
\mathrm{eff}}$ is imaginary is shown in figure 3.

Described interaction of particles and antiparticles correlates with values
of the energy-momentum tensor and other hydrodynamic quantities, calculated
in the space-time region $x^{2},t^{2}\le \Delta ^{-1}$, where the two clouds
overlap. These quantities are calculated to within $\Delta ^{0}$ 
\begin{equation}
\begin{array}{lll}
j_{0}=0 & \kappa _{0}=-E\tan \left( Et\right) & T^{00}=2E^{2}C\sin
^{2}\left( Et\right) \\ 
j_{1}=-2pC\cos ^{2}\left( Et\right) & \kappa _{1}=0 & T^{01}=0 \\ 
\rho =2C\cos ^{2}\left( Et\right) & m_{\mathrm{eff}}^{2}=-p^{2} & 
T^{11}=2p^{2}C\cos ^{2}\left( Et\right)
\end{array}
\label{a7.20}
\end{equation}
where $C$ is determined by the relation (\ref{a7.16}).

Our investigation of solutions of the free Klein-Gordon equation is a
preliminary consideration, which is necessary for correct mathematical
formulation of the collision problem and of the problem pair production as a
special case of the collision problem. Two different interpretations of the
states $\psi _{\left\{ \varepsilon _{p},\mathbf{p}\right\} }+\psi _{\left\{
\varepsilon _{p},-\mathbf{p}\right\} }$ and $\psi _{\left\{ 1,\mathbf{p}%
\right\} }+\psi _{\left\{ -1,\mathbf{p}\right\} }$ generate two different
statements of the collision problem. The conventional interpretation based
on the wave function as a fundamental object of dynamics leads to the
S-matrix theory, where detailed description of dynamic processes inside the
collision space-time region is supposed to be impossible. For solution of
these problems the energy-momentum conservation law is used mainly.
Alternative approach based on sequential consideration of fluid dynamics and
hydrodynamic quantities leads to another statement of the collision problem.
Comparison of these two different statements of the collision problem is
possible only at the detailed presentation of the alternative statement of
the collision problem. Unfortunately, such a presentation cannot be made in
the present paper.

We restrict our consideration to the problem of two elastic balls collision,
which reminds to some extent the collision problem of the microparticle
collision. At the head-on collision of two identical elastic balls they
interchange their momenta, and the whole situation may be described as one
ball passes through another one without interaction. It is a very simple
interpretation, but it is not effective, because it can be used only for
head-on collision. If the impact parameter does not vanish, or the balls are
not identical, we are forced to introduce interaction between the balls. If
we assume that the identical balls do not interact at the head-on collision,
it is rather difficult to understand, why they begin to interact with
non-vanishing impact parameter. One needs to invent a special interaction
for explanation of non-head-on collision. Is it possible to construct such a
theory of two balls collision, which is based on the supposition that the
balls do not interact at the head-on collision? Yes it is possible, but such
a theory is founded mainly on the energy-momentum conservation law, and it
does not fit for explanation of inelastic collisions, where detailed
investigation of interaction between the balls is necessary.

The approach, when the head-on collision of two identical elastic balls is
considered to be a strong interaction of two balls, which leads to an
interchange of momenta, seems to be more reasonable, because it explains
freely, why result of collision is not reduced to the interchange of momenta
at the collision with non-vanishing impact parameter. At such an approach
the interchange of momenta is explained by balanced interaction at the
head-on collision. If the impact parameter does not vanish, this balance is
violated. Although explanation of the balanced interaction is more
complicated, but application of this interaction to the case of unbalanced
interaction appears to be simpler. Something like this we have in the case
of two alternative approaches to the microparticle collision problem.

\section{Concluding remarks}

Thus, quantum system is a special case of stochastic system, and quantum
phenomena can be considered to be a result of some quantum stochasticity,
generated by the space-time properties \cite{R91}. Pair production is a
natural result of the sequential relativistic description of this
stochasticity. The statistical description of quantum phenomena generates
hydrodynamic interpretation of quantum phenomena, which does not coincide,
in general, with the interpretation, based on application of quantum
principles, and the interpretation appears to be essential for statement of
the microparticle collision problem.


\begin{thebibliography}{99}
\bibitem{R91}  Yu.A. Rylov, ''Non-Riemannian model of space-time,
responsible for quantum effects'' . \textit{J.Math. Phys}. \textbf{32},
2092-2098, (1991).

\bibitem{R2002}  Yu. A. Rylov, ''Dynamics of stochastic systems and
peculiarities of measurements in them''. (In preparation, available at
http:// arXiv.org /abs/physics /0210003).

\bibitem{R99}  Rylov Yu.A. ''Spin and wave function as attributes of ideal
fluid''. \textit{J. Math. Phys.} \textbf{40}, No.1, 256-278, (1999).

\bibitem{M26}  E. Madelung, \textit{Z.Phys.} \textbf{40} (1926) 322.

\bibitem{B26}  L. de Broglie, \textit{Comptes Rendus} \textbf{183}, 447
(1926).

\bibitem{B52}  D. Bohm, \textit{Phys.Rev.} \textbf{85,} 166, 180, (1952).

\bibitem{T52}  T. Takabayasi, \textit{Progr. Theor. Phys.}, \textbf{8}, 143,
(1952).

\bibitem{T53}  T. Takabayasi, \textit{Progr. Theor. Phys.}, \textbf{9}, 187,
(1953).

\bibitem{JZ63}  L. Janosssy and M.Ziegler, \textit{Acta Phys. Hung.}, 
\textbf{16}, 37, (1963).

\bibitem{JZ64}  L. Janosssy and M.Ziegler, \textit{Acta Phys. Hung}., 
\textbf{16}, 345, (1964).

\bibitem{HZ69}  M. Husar and M. Ziegler-Naray, \textit{Acta Phys. Acad.
Scien. Hung}., \textbf{26}, 223, (1969).

\bibitem{B73}  F.J. Belinfante, \textit{A Survey of Hidden-Variables Theories%
}, (Pergamon, Oxford, 1973) and references therein.

\bibitem{BH89}  D. Bohm and B.J. Hiley, \textit{Phys. Rep.} \textbf{172}
(1989) 93 and references therein.

\bibitem{H93}  P. Holland, \textit{The Quantum Theory of Motion}, (Cambridge
University Press, Cambridge, 1993) and references therein.

\bibitem{R89}  Yu.A. Rylov, ''The equation for isentropic motion of inviscid
fluid in terms of wave function'' \textit{J. Math. Phys.} \textbf{30},
2516-2520, (1989).

\bibitem{R95}  Rylov Yu.A. ''Pauli's electron as a dynamic system'' \textit{%
Found. Phys.} \textbf{25}, No.7, 1055-1086, (1995).

\bibitem{R995}  Rylov Yu.A. ''Dirac equation in terms of hydrodynamic
variables'' , \textit{Advances Appl. Clifford Algebras.} \textbf{5}, No. 1,
1-40, (1995).

\bibitem{L63}  C.C. Lin, \textit{Proc. International School of Physics
''Enrico Fermi''.} Course XXI, Liquid Helium , New York, Academic. 1963, pp.
93-146.

\bibitem{C57}  A. Clebsch, \textit{J. reine angew. Math.} \textbf{54 }, 293,
(1857).

\bibitem{C59}  A. Clebsch, \textit{J. reine angew. Math.} \textbf{56 }, 1,
(1859).

\bibitem{R999}  Rylov Yu.A. ''Integration of complete system of dynamic
equations for ideal fluid''. (In preparation, available at
http://arXiv.org/abs/\textit{physics/9905044)}.

\bibitem{R02}  Rylov Yu. A. ''Dynamics of stochastic systems and
peculiarities of measurements in them''. (In preparation, available at
http:// arXiv.org/abs /physics/0210003).

\bibitem{R002}  Rylov Yu.A. ''Hamilton variational principle for statistical
ensemble of deterministic systems and its application for ensemble of
stochastic systems.'' \textit{RJMP}, \textbf{9}, iss. 3, 335-348, (2002).

\bibitem{R98}  Rylov Yu.A. ''Quantum mechanics as a dynamic construction''. 
\textit{Found. Phys.} \textbf{28}, No.2, 245-271, (1998).

\bibitem{R94}  Rylov Yu.A., ''Statistical conception of quantum field
theory''. \textit{J. Math. Phys.} \textbf{35}, 3922-3935, (1994).

\bibitem{R001}  Rylov Yu.A. Pair production problem and canonical
quantization of nonlinear scalar field in terms of world lines.'' (In
preparation, available at http://arXiv.org/abs /\textit{hep-th /0106169).}

\bibitem{GJ68}  J. Glimm and A.\ Jaffe, \textit{Phys. Rev. }\textbf{176}
(1968) 1945.

\bibitem{GJ70}  J. Glimm and A.\ Jaffe, \textit{Ann. Math. }\textbf{91}
(1970) 362.

\bibitem{GJ970}  J. Glimm and A.\ Jaffe, \textit{Acta Math. }\textbf{125}
(1970) 203.

\bibitem{GJ72}  J. Glimm and A.\ Jaffe, \textit{J. Math. Phys. }\textbf{13}
(1972) 1568.

\bibitem{FV58}  Feshbach H., Villars F., Rev. Mod. Phys. \textbf{30}, 24,
(1958).
\end{thebibliography}
\end{document}